\documentclass[12pt]{article}

\usepackage{graphicx,epsf,subfigure}
\usepackage{amsmath,amsfonts,natbib,verbatim,booktabs,longtable,multirow,lscape}
\usepackage{enumerate, color, url}
\usepackage{dsfont,ctable}
\usepackage[margin = 1in]{geometry}
\usepackage{setspace}
\usepackage{algorithm}
\usepackage{wasysym}
\usepackage{makecell}

\newcommand{\summ}{\sum_{m=1}^{M}}

\newcommand{\dA}{\mathds{A}}
\newcommand{\dP}{\mathds{P}}

\newcommand{\bB}{\mathbf{B}}

\newcommand{\bI}{\mathbf{I}}

\newcommand{\bG}{\mathbf{G}}
\newcommand{\bH}{\mathbf{H}}

\newcommand{\bX}{\mathbf{X}}
\newcommand{\bZ}{\mathbf{Z}}

\newcommand{\bb}{\mathbf{b}}

\newcommand{\bzero}{\mathbf{0}}

\newcommand{\btheta}{\boldsymbol{\theta}}

\newcommand{\bbeta}{\boldsymbol{\beta}}
\newcommand{\bxi}{\boldsymbol{\xi}}

\newcommand{\bgamma}{\boldsymbol{\gamma}}
\newcommand{\bnu}{\boldsymbol{\nu}}

\newcommand{\calA}{\mathcal{A}}

\newcommand{\Prob}{\mbox{Pr}}
\newcommand{\ts}{\textrm{ts}}
\newcommand{\tr}{\textrm{tr}}

\title{Bayesian Semiparametric Estimation of Cancer-specific Age-at-onset Penetrance with Application to Li-Fraumeni Syndrome}
\author{Seung Jun Shin$^1$,  Ying Yuan$^{2,*}$, Louise C. Strong$^2$, \\ Jasmina Bojadzieva$^2$, and Wenyi Wang$^{2,*}$ \\ \smallskip
\normalsize{ \it Korea University$^1$ and The University of Texas MD Anderson Cancer Center$^2$}\\ \bigskip
\normalsize $*$ Authors to correspondence}
\date{}

\begin{document}

\maketitle

\begin{abstract}
Penetrance, which plays a key role in genetic research, is defined as the proportion of individuals with the genetic variants (i.e., {genotype}) that cause a particular trait and who have clinical symptoms of the trait (i.e., {phenotype}). We propose a Bayesian semiparametric approach to estimate the cancer-specific age-at-onset penetrance in the presence of the competing risk of multiple cancers. We employ a Bayesian semiparametric competing risk model to model the duration until individuals in a high-risk group develop different cancers, and accommodate family data using family-wise likelihoods. We tackle the ascertainment bias arising when family data are collected through probands in a high-risk population in which disease cases are more likely to be observed. We apply the proposed method to a cohort of 186 families with Li-Fraumeni syndrome identified through probands with sarcoma treated at MD Anderson Cancer Center from 1944 to 1982.

\vspace{4mm}
\end{abstract}

\section{Introduction} \label{s::intro}
The Li-Fraumeni syndrome (LFS) is a rare disorder that substantially increases the risk of developing several cancer types, particularly in children and young adults. It is characterized by autosomal dominant mutation inheritance with frequent occurrence of several cancer types: soft tissue/bone sarcoma, breast cancer, lung cancer, and other types of cancer that are grouped together as ``other cancers" \citep{nichols2001germ,birch2001relative}. A majority of LFS is caused by germline mutations in the TP53 tumor suppressor gene \citep{malkin1990germ,srivastava1990germ}.

The LFS data that motivate our work are family data collected through patients diagnosed with pediatric sarcoma (i.e., probands) who were treated at MD Anderson Cancer Center from 1944 to 1982 and their extended kindred. The data consist of 186 families, with a total of 3686 subjects. The size of the families ranges from 3 to 717, with the median at 7. This dataset is the longest followed-up cohort in the world (followed up for 20-50 years), and among the largest collection of TP53 mutation carriers in all cohorts that are available for LFS. Considering the prevalence of TP53 mutations in a general population is as low as 0.0001 to 0.003, this dataset provides a specially enriched collection of TP53 mutations, which then allow us to characterize its effect on a diverse cancer outcomes. For each subject, the duration until he/she develops cancer is recorded as the primary endpoint. Although it is possible for LFS patients to experience multiple cancers during their lifetime, here, we focus on only the time to the first primary cancer since only a few patients represented in the database experienced multiple primary cancers. Table \ref{tb::summary2} shows the cancer-specific summaries for the LFS data. Further descriptions of the data are provided by \citet{lustbader1992segregation}, \citet{strong1992li}, and \citet{hwang2003germline}.

\begin{table} [b]
\caption{Frequency table for LFS data.
} \label{tb::summary2}
\centering
\begin{tabular}{l ccccc r}
\hline
 Gender & Genotype        & Breast &  Sarcoma & Others& Censored & Total \\ \hline
        & Unknown         &      0 &       11 & 	 130&   1275	&	1416   \\
 Male   & Wildtype        &      0 &       76 &	  30&    295	&	 401   \\
        & Mutation        &      0 &       12 &	  27&      9	&	  48   \\ \cline{2-7}
        & Subtotal        &      0 &       99 &    187 &   1579	&   1865   \\ \hline
        & Unknown         &     39 &        4 &	  62 &   1204	&	1309   \\
 Female & Wildtype        &      8 &       96 &	  17 &    343	&	 464   \\
        & Mutation        &     19 &       12 &	   7 &     10	&	  48   \\ \cline{2-7}
        & Subtotal        &     66 &      112 &	  86 &   1557	&   1821   \\ \hline
\multicolumn{2}{c}{Total} &     66 &      211 &	 273 &   3136	&   3686   \\ \hline
\end{tabular}
\end{table}

The primary objective here is to estimate the cancer-specific age-at-onset penetrance as a measure of the risk of experiencing a specific cancer for a person with a specific genotype (i.e., TP53 mutation status). Penetrance, which plays a crucial role in genetic research, is defined as the proportion of individuals with the genetic variants (i.e., {genotype}) that cause a particular trait who also have clinical symptoms of the trait (i.e., {phenotype}).
When the clinical traits of interest are age-dependent (e.g., cancers), it is often more desirable to estimate the age-at-onset penetrance, defined as the probability of disease onset by a certain age, while adjusting for additional covariates if necessary.
For the LFS study, the age-at-onset penetrance is defined as the conditional probability of having LFS-related cancers by a certain age given a certain TP53 mutation status.  Cox proportional hazard regression models \citep[][among many others]{GaudermanFaucett1997, WuStrongShete2010} have been most widely used for this task. Other approaches have included nonparametric estimation \citet{wang2007nonparametric} and parametric estimation based on logistic regression \citep{abel1990time} or a Weibull model \citep{hashemian2009kin}.

Estimating the age-at-onset penetrance for the LFS data is challenging  for several reasons. 
First, LFS involves multiple types of cancer, and subjects have simultaneous competing risks of developing multiple types of cancer. \citet{chatterjee2003adjustment} proposed a penetrance estimation method under a competing risk framework for a kin-cohort design. However, their method is not directly applicable if the pedigree size is large and/or there is additional genetic information from relatives. \citet{gorfine2011frailty} and \citet{gorfine2014calibrated} proposed frailty-based competing risk models for family data, assuming that genotypes are completely observed for all family members, which is not the case for the LFS data.

Second, the genotype (i.e., TP53 mutation status) is not measured for the majority (about 74\%) of subjects and the LFS data are clustered in the form of families. 
Accommodating the missing data and accounting for family or pedigree data structure are statistically and computationally challenging. As shown later, to efficiently utilize the observed genotype data nested in the family structure, we need to marginalize the likelihood over (or integrate out) all possible genotypes for subjects with missing genotype information, and meanwhile take into account the available genotypes in the family under the given pedigree structure.

Third, the LFS data are not a random sample, but have been collected through probands diagnosed with sarcoma at young ages. That is, the data oversampled sarcoma patients. Such a sampling scheme inevitably creates bias, known as \textit{ascertainment bias}, and should be properly adjusted to obtain unbiased results. Several likelihood-calibrated models have been developed to correct the ascertainment bias, including the retrospective model \citep{KraftThomas2000}, the conditional-on-ascertainment variable model \citep{ewens1986resolution,pfeiffer2001inference}, and the ascertainment-corrected joint model \citep{KraftThomas2000,IversenChen2005}, among others.

To address these challenges, in  this article, we develop a Bayesian semiparametric approach to estimate the cancer-specific age-at-onset penetrance in the presence of the competing risks of developing multiple cancers. We employ a Bayesian semiparametric competing risk model to model the time to different types of cancer and introduce the family-wise likelihood to minimize information loss from missing genotypes and harness the information contained in the pedigree structure. We employ the peeling algorithm \citep{elston1971general} to evaluate the family-wise likelihood, and utilize the ascertainment-corrected joint model (Kraft and Thomas, 2000) to correct the ascertainment bias.

The rest of the article is organized as follows. In Section \ref{s::model}, we define the cancer-specific age-at-onset penetrance and describe our Bayesian semiparametric competing risk model including details about the family-wise likelihood and the ascertainment bias correction. In Section \ref{s::sampling}, we provide an algorithm to fit the models and carry out a simulation study in Section \ref{s::simulation}. We apply the proposed methodology to the LFS data in Section \ref{s::analysis}.  Discussions follow in Section \ref{s::discussion}.

\section{Model} \label{s::model}
\subsection{Cancer-specific Age-at-onset Penetrance}  \label{ss::cs.penetrance} 
Let $G$ denote a subject's genotype, and $X$ denote the baseline covariates (e.g., gender). Suppose that $K$ types of cancer are under consideration and compete against each other such that the occurrence of one type of cancer censors the other types of cancer.  Let $T_k$ denote the time to the $k$th type of cancer, $k=1, \ldots, K$, and define $T = \min_{k \in \{1, \cdots, K\}} T_k$ and $Y = \min \{T, C \}$, where $C$ is a conditional random censoring time given $G$ and $X$, i.e., $T \bot C|G,X$.
Let $D$ denote the cancer type indicator, with $D=k$ if $T=T_k<C$ (i.e., the $k$th type of cancer that occurs); otherwise, $D=0$ (i.e., censored observation).
The actual observed time-to-event data are $H=(Y, D)$.

Traditionally, when analyzing subjects at risk of developing a single disease, the age-at-onset penetrance is defined as the probability of having the disease at a certain age given a certain genotype.  In order to study LFS, where subjects simultaneously have the (competing) risk of developing multiple types of cancer,  this standard definition must be extended. Borrowing ideas from the competing risk literature, we define the $k$th \textit{cancer-specific age-at-onset penetrance}, denoted by $q_k(t|G, X)$,  as the probability of having the $k$th type of cancer  at age $t$ prior to developing other cancers (competing risks), given a specific genotype $G$ and additional baseline covariates $X$ if necessary, that is,
\begin{align} \label{eq::cs.penetrance}
q_k(t|G,X) = {\rm Pr}(T \le t, D = k|G, X), \qquad  k = 1, \cdots, K.
\end{align}
The cancer-specific penetrance $q_k(t|G,X)$ can be estimated as
\begin{equation} \label{eq::cs.penetrance2}
q_k(t|G,X) = \int_{0}^{t}\lambda_k(u|G, X) S(u|G, X) du, \qquad  k = 1, \cdots, K,
\end{equation}
where
\begin{equation} \label{eq::cs.hazard}
\lambda_k(t|G, X) = \lim_{h \downarrow 0} \frac{\Prob(t \le T < t+h, D = k | T > t, G, X)}{h},
\end{equation}
and
$$S(t|G, X) = \exp\left\{-\sum_{k=1}^K \Lambda_k(t|G, X)\right\},$$ with $\Lambda_k(t|G, X) = \int_{0}^{t}\lambda_k(u|G,X) du$. In the competing risk literature, $\lambda_k(t|G, X)$ and $\Lambda_k(t|G, X)$ are referred to as the cancer-specific hazard and cancer-specific cumulative hazard, respectively. We note that it may be tempting to define the cancer-specific age-at-onset penetrance function as ${\rm Pr}(T_{k} \le t | G, X)$, which is analogous to the conventional definition of penetrance for a single disease. However, that quantity is not identifiable in nonparametric models \citep{tsiatis1975nonidentifiability}.

Besides cancer-specific penetrance, it is often of practical interest to estimate the overall age-at-onset penetrance,  defined as
\begin{align} \label{eq::penetrance}
q(t|G, X) = \Prob(T \le  t | G, X),
\end{align}
which is the probability that a subject has any type of cancer by age $t$ given his/her genotype $G$ and baseline characteristics $X$, and can
 be calculated through the cancer-specific penetrance $q_k(t|G, x)$ using $q(t|G, X) = \sum_{k=1}^K q_k(t|G, X)$.

\subsection{Competing Risk Model}  \label{ss::model} 

In the rest of the article, without loss of generality, we focus on the LFS data where $X$ denotes gender coded as 1 for the male and 0 for the female, and $G$ denotes TP53 mutation status. As LFS is autosomal dominant, let $G=1$ denote genotype $ Aa$ or $AA$, and $G=0$ denote genotype $aa$, where $A$ and $a$ denote the (minor) mutated and wildtype alleles in the TP53 tumor suppressor gene, respectively. Let $\bZ = (G, X, G \times X )^T$, with $G \times X$ denoting the interaction between $G$ and $X$. 
We model the hazard for the $k$th type of cancer, say $\lambda_k(t|\bZ, \xi_{i, k})$, using a frailty model as follows:
\begin{align} \label{eq::frailty.model}
\lambda_k(t|\bZ, \xi_{i, k}) = \lambda_{0,k} (t) \xi_{i,k} \exp\{\bbeta_{k}^T \bZ\}, \qquad k = 1, \cdots, K,
\end{align}
where  $\bbeta_k$ denotes the regression coefficient parameter vector; $\lambda_{0,k}(t)$ is a baseline hazard function; and $\xi_{i,k}$ is the $i$th family-specific frailty (or random effect) used to account for the within-family correlation induced by non-genetic factors that are not included in $X$. The pedigree information (or genetic relationship) within a family will be incorporated through the family-wise likelihood described in Section 2.3.  We assume that $\xi_{i,k}$ follows a gamma distribution, $\xi_{1,k}, \cdots, \xi_{I,k} \sim Gamma(\nu_k, \nu_k)$. Such a gamma frailty has been widely used in frailty models \citep{duchateau2007frailty}.

Under this model, the cancer-specific age-at-onset penetrance can be expressed as
\begin{align}
q_k(t|\bZ) &=  \int_0^t \int_{\bxi \in [0, \infty)^K} \lambda_k (u|\bZ, \xi_k) S(u|\bZ,\bxi) f(\bxi | \bnu) d\bxi du \nonumber \\
&=\int_0^t \frac{\nu_k}{(\nu_k - \log \{S_k^*(u|\bZ)\})} \lambda_{0,k}(u) \exp\{\bbeta_k^T\bZ\} S(u|\bZ) du, \label{eq::cs.penetrance3}
\end{align}
where $S_k^*(t|\bZ) = \exp\left\{-\int_{0}^t \lambda_{k,0}(u) \exp\{\bbeta_k^T \bZ\} du \right\}$ and $S(t|\bZ) = \prod_{k=1}^K S_k(t|\bZ)$ with
\begin{align*}
S_k(t|\bZ) 
& =  \int_{0}^{\infty} \exp\left\{-\int_{0}^t \lambda_k (u|\bZ,\xi_k) du \right\} f(\xi_k | \nu_k) d \xi_k \\
& = \left[ \frac{\nu_k}{\nu_k - \log\{S_k^*(t|\bZ)\}} \right]^{\nu_k}.
\end{align*}

Because the penetrance depends on the survival function, it is imperative to specify the baseline hazard  $\lambda_{0,k} (t)$, which appears in (\ref{eq::frailty.model}). To this end, we propose to approximate the cumulative baseline hazard $\Lambda_{0,k} (t) = \int_{0}^{t} \lambda_{0,k} (s) ds$ via Bernstein polynomials \citep{lorentz1953bernstein} since $\Lambda_{0,k} (t)$ is monotone increasing. Bernstein polynomials are popular in Bayesian nonparametric function estimation, with shape restrictions due to desired properties such as the optimal shape restriction property \citep{carnicer1993shape} and the convergence property of their derivatives \citep{lorentz1953bernstein}.
Without loss of generality, we assume $t$ has been rescaled, e.g., by the largest observed time, such that $t \in [0,1]$. 
Now, we have $\Lambda_{0,k}(t)$ 
approximated by Bernstein polynomials of degree $M$ as follows \citep{chang2005bayesian}.
\begin{eqnarray}
\Lambda_{0,k}(t) \approx \sum_{l=1}^{M} \omega_{l,k} {M \choose l} t^{l} (1-t)^{M-l},  \label{eq::beta.approximation}
\end{eqnarray}
where $\omega_{l,k} = \Lambda_{0,k}(l/M)$ and $\omega_{1,k} \le \cdots \le \omega_{M, k}$ to ensure that $\Lambda_{0,k}(t)$ is monotone increasing. Notice that $l$ is running from 1 because of $\Lambda_{0,k}(0) = 0$. Applying the re-parameterization of $\gamma_{l,k} = \omega_{l,k} - \omega_{l-1,k}$ with $\omega_{0, k} = 0$ and $\gamma_{l,k} \ge 0, l = 1, \cdots, M$, the right-hand side of (\ref{eq::beta.approximation}) can be equivalently rewritten as 
\begin{align}
\summ \gamma_{m,k} \int_0^{t} \frac{u^{m} (1-u)^M-m}{Beta(m, M-m+1)} du = \bgamma_k^T \bB_M(t), \label{eq::beta.approximation2}
\end{align}
where $\bB_M(t) = (B_M(t,1), \cdots, B_M(t,M))^T$, with $B_M(t,m)$ being the distribution function of the beta distribution evaluated at the value of $t$ with parameters $m$ and $M-m+1$, and $\bgamma_k = (\gamma_{1,k}, \cdots, \gamma_{M,k})^T$ \citep{mckay2011variable}. 
Therefore,  it follows that
\begin{eqnarray} \label{eq::baseline}
\lambda_{0,k}(t) \approx \bgamma_k^T \bb_M(t),
\end{eqnarray}
where $\bb_M(t) = (b_M(t,1), \cdots, b_M(t,M))^T$ and $b_M(t, m) = \partial B_M(t,m)/\partial t$ (i.e., associated beta density). Finally, the frailty model (\ref{eq::frailty.model}) can be written as
\begin{align} \label{eq::cs.penetrance4}
\lambda_k(t|\bZ; \bbeta_k, \bgamma_k, \xi_{i,k}) = \{\bgamma_k^T \bb_M(t)\} \xi_{i,k} \exp\{\bbeta_{k}^T \bZ\}.
\end{align}

The proposed nonparametric baseline hazard model \eqref{eq::baseline} is more flexible than parametric models, such as exponential and Weibull models, without imposing a restrictive parametric structure on the shape of the baseline hazard. Compared to the piecewise constant hazard model, our approach produces a smooth estimate of hazard and also avoids selection of knots, which is often subjective. The numerical comparison of different baseline models is provided in Section \ref{ss:compare} and {\textit{Supplementary Materials}} Section C.

\subsection{Family-wise Likelihood} 

Let $i$ index the family and $j$ index the individual within the family, where $i=1, \cdots, I$, and  $j = 1, \cdots, n_i$. For the $i$th family,  let $\bH_{i} = (H_{i1}, \cdots, H_{i n_i})$ with $H_{ij} = (Y_{ij}, D_{ij})$ and $\bX_{i} = (X_{i1}, \cdots, X_{i n_i})$. Let $\bG_{i, obs}$ and $\bG_{i, mis}$ respectively denote the observed and missing parts of genotype data $\bG_{i} = (G_{i1}, \cdots, G_{i n_i})$, i.e., $\bG_{i}=(\bG_{i, obs}, \bG_{i, mis})$ if we ignore the order of the elements for simplicity. Conditional on frailty $\bxi_{i}=(\xi_{i,1}, \cdots, \xi_{i,K})$, the likelihood of $\bH_{i}$ for the $i$th family is $\Pr (\bH_i | \bG_{i, obs}, \bX_i, \btheta , \bxi_{i})$ which we call the family-wise likelihood, where $\btheta = \{(\bbeta_k^T, \bgamma^T_k) : k = 1, \cdots, K \}$ denotes a vector of model parameters except the frailty.  

Evaluation of the family-wise likelihood $\Pr (\bH_i | \bG_{i, obs}, \bX_i, \btheta , \bxi_{i})$ is not trivial because the individual disease histories $H_{i1}, \cdots, H_{in_i}$ are not conditionally independent given $\bG_{i, obs}$ and $\bxi_{i}$, due to the dependency through $\bG_{i,mis}$. Note that  $H_{i1}, \cdots, H_{in_i}$ will be conditionally independent when conditional on complete genotype data $\bG_{i}$ and $\bxi_{i}$. In this article, we use Elston-Stewart's peeling algorithm \citep{elston1971general,lange1975extensions,FernandoStrickerElston1993} to compute the family-wise likelihood, described as follows. We assume that there is no loop in the pedigree, which is generally true in practice, and suppress the family subscript $i$ and the conditional arguments except $\bG_{obs}$ for notational brevity.

A pedigree without loop can be partitioned into two disjoint groups, known as anterior and posterior,  that are connected only through an arbitrary pivot member, say $j$. The anterior are the member in the pedigree who are connected to the pivot member through his/her parents, and the posterior are the member in the pedigree who are connected to the pivot member through his/her spouse and offsprings, see Figure \ref{fg:ped.sim} for an example. In our implementation, we use the proband as the pivot member of each family. Let  $\bH_j^-$, and $\bH_j^+$ denote the phenotypes of anterior and posterior, respectively. We partition $\bH = (\bH_j^-, H_j, \bH_j^+)$. Because anterior and posterior are connected only through the pivot member $j$,  $\bH_j^-$ and $\bH_j^+$ are conditionally independent given pivot member's genotype $G_j$.

If $G_j$ is unobserved, the family-wise likelihood $P(\bH|\bG_{obs})$ can be written as 
\begin{align} 
\Pr(\bH|\bG_{obs}) 
& = \sum_{G_{j}} {\Big\{}\Pr(G_j | \bG_{obs}) \Pr(\bH_j^-, H_j, \bH_j^+ | G_{j}, \bG_{obs}) {\Big\}}  \nonumber  \\
& = \sum_{G_{j}} \Big\{\Pr(G_{j}|\bG_{obs}) \Pr(\bH_j^- | G_{j}, \bG_{obs}) \Pr(H_j | G_{j}, \bG_{obs}) \Pr(\bH_j^+ | G_{j}, \bG_{obs}) \Big\}\nonumber \\
& = \sum_{G_{j}} {\Big\{}\dA_j(G_{j}|\bG_{obs})  \Pr(H_j|G_{j})  \dP_j(G_j|\bG_{obs}) {\Big\},} \label{eq:f.lk}
\end{align}
where 
\begin{align*}
\begin{array} {ll c l}
\mbox{(Anterior probability of $j$)}  & \dA_j(G_j|\bG_{obs}) & = & \Pr(\bH_{j}^-, G_{j}|\bG_{obs}), \\
\mbox{(Posterior probability of $j$)} & \dP_j(G_j|\bG_{obs}) & = & \Pr(\bH_{j}^+| G_{j},\bG_{obs}),
\end{array}
\end{align*}
and the individual likelihood $\Pr (H_j | G_j)$ is computed from the proposed model as
\begin{align*}
\Pr (H_j | G_j) : = \Pr (H_{ij} | G_{ij}, X_{ij}, \btheta , \bxi_i) \propto \prod_{k = 1}^K \left\{\lambda_{k}(Y_{ij}|\bZ_{ij}, \btheta, \bxi_i)\right\}^{\Delta_{ijk}}
\exp \left\{ - \Lambda_k(Y_{ij}|\bZ_{ij}, \btheta, \bxi)\right\},
\end{align*}
with $\Delta_{ijk} = 1$ if $D_{ij} = k$ and 0 otherwise \citep{prentice1978analysis,maller2002analysis}. 
In the case that $G_j$ is observed, the summation in \eqref{eq:f.lk} is not needed and the family-wise likelihood is reduced to
\begin{align} 
\Pr(\bH|\bG_{obs}) 
& = \dA_j(G_{j}|\bG_{obs})  \Pr(H_j|G_{j})  \dP_j(G_j|\bG_{obs}). \label{eq:f.lk2}
\end{align}
To calculate $\dA_j(G_j|\bG_{obs})$ and $\dP_j(G_j|\bG_{obs})$,  $\bH_j^-$ and $\bH_j^+$ can be further partitioned into anterior and posterior in a similar way as above. Thus, the family-wise likelihood $\Pr(\bH|\bG_{obs})$ can be evaluated in a recursive way. An illustrative example of using the peeling algorithm to evaluate the family-wise likelihood is provided in \textit{Supplementary Materials} Section A.  \citet{FernandoStrickerElston1993} provides the details on the recursive formulation of the algorithm.

\subsection{Ascertainment Bias Correction} \label{ss::ascertainment} 
For studies of rare diseases, such as LFS, ascertainment bias is inevitable when family data are collected through probands in high-risk populations in which disease cases are more likely to be observed. In order to correct the ascertainment bias, we employ the ascertainment-corrected joint (ACJ) likelihood \citep{KraftThomas2000,IversenChen2005}. In particular, we closely follow the approach proposed by \citet{IversenChen2005}. 
Let ${\mathcal A}_i$ denote the ascertainment indicator variable, such that ${\mathcal A}_i=1$ if the $i$th family is ascertained and $0$ otherwise. In the LFS data, a family is ascertained and included in the sample only if the proband is diagnosed with sarcoma. Following the idea of \citet{IversenChen2005}, the ACJ likelihood for the LFS data is given by  
\begin{align} \label{eq::abc}
& \Prob(\bH_i, \bG_{i, obs}| \bX_i, \btheta, \bxi_i, \calA_i=1) \nonumber \\
& \qquad \qquad 
= \frac{\Pr(\calA_i=1|\bH_i, \bG_{i, obs}, \bX_i, \btheta, \bxi_i)
\Pr(\bH_i| \bG_{i, obs}, \bX_i, \btheta, \bxi_i)\Pr(\bG_{i, obs}| \bX_i, \btheta, \bxi_i)}{\Pr(\calA_i=1| \bX_i, \btheta, \bxi_i)}.
\end{align}
Because the ascertainment decision is made on the basis of $H_{i1}$ (i.e., phenotype of the proband) in a deterministic way, the first term in the numerator of equation \eqref{eq::abc}, i.e., $\Pr(\calA_i=1|\bH_i, \bG_{i, obs}, \bX_i, \btheta, \bxi_i)$,  is independent of the model parameters $\btheta$ and $\bxi_i$. In addition, the third term in the numerator of equation \eqref{eq::abc}, i.e., $\Pr(\bG_{i, obs}| \bX_i, \btheta, \bxi_i)$ 
is also independent of both $\btheta$ and $\bxi_i$, the parameters of the penetrance model. As a result, we have
\begin{equation} \label{eq::acj}
\Prob(\bH_i, \bG_{i, obs}, \bX_i| \btheta, \bxi_i, {\mathcal A}_i=1)
\propto \frac{\Prob(\bH_i | \bX_i, \bG_{i, obs}, \btheta, \bxi_i)}{\Prob({\mathcal A}_i =1 |\bX_i, \btheta, \bxi_i)}.
\end{equation}
This means that the ascertainment bias can be corrected by inverse-probability weighting the likelihood by the corresponding ascertainment probability, which is given by
\begin{align} \label{eq::abc.lfs}
\Prob({\mathcal A}_i=1| \bX_i, \btheta, \bxi_i)  
= \sum_{H_{i1}} \Prob({\mathcal A}_i=1|H_{i1})  \Prob(H_{i1}|X_{i1}, \btheta, \bxi_i).
\end{align}
In the LFS data, 
a family is ascertained only if the proband is diagnosed with sarcoma (coded as $D=2$). We assume that the sarcoma patients visiting MD Anderson Cancer Center are not be very different from the ones visiting other clinics, then it follows
$$
\Prob({\mathcal A}_i=1|Y_{i1}, D_{i1} = 2) = 1 \,\,{\rm and} \,\, \Prob({\mathcal A}_i=1|Y_{i1}, D_{i1} \neq 2) = 0.
$$  
The assumption is empirically acceptable in our application, and partially validated in Section 5.4 by comparing our estimates for non-carriers to those from the US population. In general, however, the assumption may not be valid, then we cannot generalize our results to the US population, but to the patients visiting MD Anderson Cancer Center only.

Recalling $H_{i1}=(Y_{i1}, D_{i1})$, the ascertainment probability  \eqref{eq::abc.lfs} is given by
\begin{align}  \label{eq::abc.prob}
\Prob({\mathcal A}_i=1|\bX_i, \btheta, \bxi_i) 
& = \Prob(Y_{i1}, D_{i1} = 2 |X_{i1}, \btheta, \bxi_i) \notag \\
& = \sum_{G}
\Prob(Y_{i1}, D_{i1} = 2 |G, X_{i1}, \btheta, \bxi_i) \Prob(G|X_{i1}) \notag \\
& = \sum_{G} \Bigg[
\lambda_{2}(Y_{i1}|G, X_{i1}, G \times X_{i1}, \bbeta_2, \bgamma_2, \xi_{i,2}) \nonumber \\
& \qquad \times  \exp \bigg\{ - \sum_{k = 1}^K  \Lambda_k(Y_{i1}|G, X_{i1}, G \times X_{i1}, \bbeta_k, \bgamma_k, \xi_{i,k})\bigg\} \Prob(G|X_{i1})\Bigg].
\end{align}
The gender-specific prevalence $\Prob(G|X_{i1})$ is often assumed to be gvien 
when estimating the penetrance \citep{IversenChen2005}. In our application, the TP53 mutation prevalence is independent of gender i.e., $\Pr(G|X) = \Pr(G)$, and it can be calculated on the basis of the mutated allele frequency $\phi_A$, i.e., $\Prob(G = 0)=(1-\phi_A)^2$ and $\Prob(G = 1)=1-(1-\phi_A)^2$.  
The prevalence of a germline TP53 mutation in the Western population is known to be $\phi_A=0.0006$ \citep{lalloo2003prediction}.

As shown above, the key is that we assume that the mutated allele frequency $\phi_A$ is known or can be reliably estimated from external data sources. Given a known mutated allele frequency $\phi_A$, the frequency of each genotype $G$ can be determined using the Mendelian laws of inheritance. Thus, coupling with the penetrance model, the sampling probability can be estimated, e.g., equation (\ref{eq::abc.prob}), and used to inversely weight the observed data likelihood to make inference for the target population. For many genetic studies, it is often reasonable to assume that  the mutated allele frequency $\phi_A$ is known or can be reliably estimated from external data sources. 

The ACJ likelihood of the entire data of $I$ mutually independent families is given by the product of \eqref{eq::acj} 
\begin{align*} 
\Pr (\bH, \bG_{obs}|\bX, \btheta, \bxi, {\mathcal{A}}) \propto \prod_{i=1}^I \frac{\Pr (\bH_i | \bG_{i, obs}, \bX_i, \btheta , \bxi_{i})}{\Prob({\mathcal A}_i=1|\btheta, \bxi_i)},
\end{align*}
where $\bH = (\bH_i, \cdots, \bH_I)$, $\bG_{obs} = (\bG_{1, obs}, \cdots, \bG_{I,obs})$ and ${\mathcal{A}} = ({\mathcal{A}}_1, \cdots, {\mathcal{A}}_I)$.

\section{Prior and Posterior Sampling} \label{s::sampling} 

We use an independent normal prior for $\bbeta_k$, i.e.,
$
\bbeta_k \sim N(\bzero, \sigma^2 \bI),
$
where $\bzero$ and $\bI$ denote a zero vector and an identity matrix, respectively, and we set a large value of $\sigma$ 
for vague priors.
For the nonnegative parameter $\gamma_{m,k}, m = 1, \cdots, M, k = 1, \cdots, K$ for the baseline hazard, we use the noninformative flat prior. We assign $\nu_k$, $k = 1, \cdots, K$,  the independent vague gamma prior $Gamma(0.01, 0.01)$. See Section \ref{ss::sensitivity} for the results of the sensitivity analysis of $\gamma_{m,k}$ and $\nu_k$. For the choice of $M$, a large value provides more flexibility to model the shape of the baseline hazard, but at the cost of increasing the computational burden. \citet{GelfandMallick1995} suggest that a small value of $M$ works well for most applications. We set $M = 5$ in the analysis.

Let $\Prob(\btheta)$ and $\Prob(\bnu)$ denote the prior distributions of $\btheta$ and $\bnu$, respectively. The joint posterior distribution of $\bnu$, $\bxi$ and $\btheta$ is given by
\begin{align*}
\Prob(\btheta, \bxi, \bnu|\bH, \bG_{obs}, \bX, {\mathcal{A}}) \propto \Pr (\bH, \bG_{obs}|\bX, \btheta, \bxi, {\mathcal{A}}) \cdot  \Prob(\btheta) \cdot \Prob(\bxi|\bnu) \cdot \Prob(\bnu).
\end{align*}

We employ the random walk Metropolis-Hastings algorithm within Gibbs sampler to sample the posterior distribution.  We generate 100,000 posterior samples in total and take every fifth sample for thinning after discarding the first 10,000 samples for burn-in. We implement the Markov chain Monte Carlo (MCMC) algorithm in \texttt{R}, which takes about three seconds per single MCMC iteration. We observe that the physical computing time is approximately linear, corresponding to the number of families, $I$, regardless of the family size $n_i$. 

\section{Simulation} \label{s::simulation}

\begin{figure}[!b]
\centering
\includegraphics[width = 0.9\textwidth]{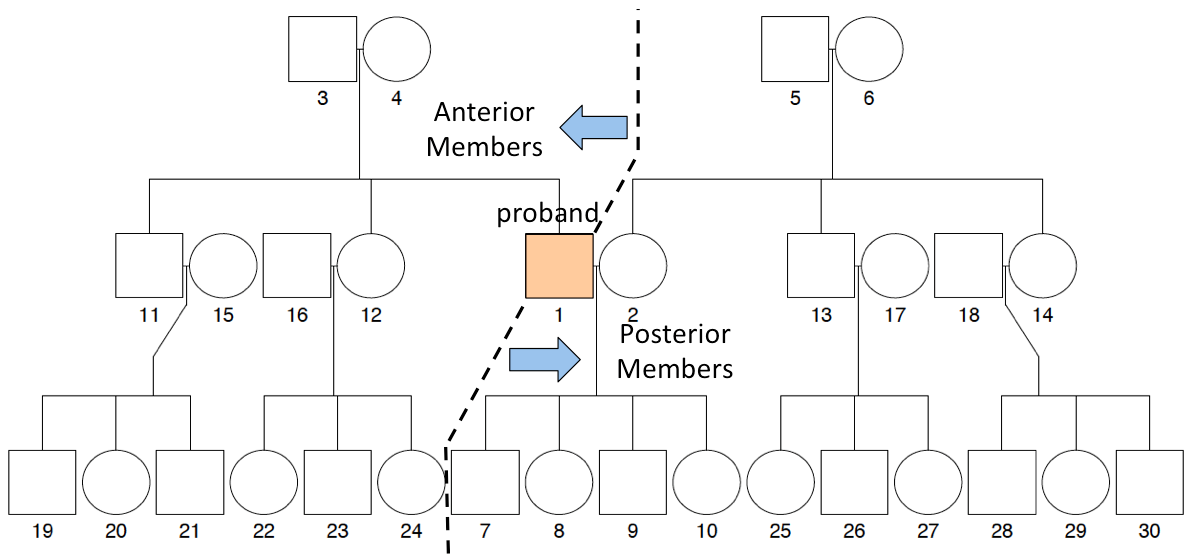}
\caption{Pedigree of the simulated family of three generations with 30 members. } \label{fg:ped.sim}
\end{figure}

We conduct a simulation study to evaluate the performance of the proposed method. Suppose that there are two competing cancers, indicated by $D=1$ and 2, respectively. We simulate 200 families of three generations with 30 members (see Figure \ref{fg:ped.sim}) that are collected through probands indexed by $\{1\}$ in Figure \ref{fg:ped.sim} with the second type of cancers (i.e., $D=2$), as follows: 
\begin{enumerate}
\item We first simulate a genotype $G \sim Bernoulli (0.0001)$ for the proband. Given $G$, we then simulate his/her true time to cancer, $T_k, k = 1,2$,  from the following cancer-specific frailty model:  
\begin{align} \label{eq:model} 
\lambda_k(t|G) = \lambda_{0,k}(t) \xi_k \exp(\beta_k G), \qquad k = 1, 2,
\end{align}
with $\beta_1 = 4, \beta_2 = 10$, $\lambda_{0,1}(t) = 0.1 , \lambda_{0,2}(t) = 0.0005$, and $\xi_1, \xi_2 \stackrel{iid}{\sim} Gamma(0.25, 0.25)$.
We choose these simulation parameters such that the second type of cancer (i.e., $D=2$) is rare with the prevalence of about 0.0003, while that of the first type of cancer is about 0.05. Random censoring time $C$ is simulated from $Exponential(2)$. To mimic the ascertainment process of the LFS data, only probands with $D=2$ are selected and included in the sample as probands. We repeat the above procedure until 200 probands are collected.

\item Given probands' data, we generate genotypes of their family members as follows. If proband $\{1\}$ is a non-carrier ($G = 0$), all family members are set as non-carriers; otherwise, we randomly select one of proband's parents $\{3, 4\}$ as a carrier and set the another as a non-carrier.  Offsprings and siblings of the proband, including $\{7,8,9,10,11,12\}$, are set as carriers with probability 0.5. 
If $\{11\}$ is carrier, his offsprings, including $\{19, 20,21\}$, are set as carries with probability 0.5, otherwise set as noncarriers. Genotypes of $\{22,23,24\}$ are generated similarly  based on the genotype of their mother $\{12\}$. 
Assuming that the mutation is extremely rare, the family members who are not genetically related with the proband, including $\{2,5,6,13,14,15,16,17,18,25,26,27,28,29,30\}$, are set as non-carriers.
\item Given the genotypes, the time to cancer of the family members are generated from model \eqref{eq:model}.  
\item Lastly, we randomly delete genotypes for a half of subjects who are not a proband.
\end{enumerate}

We set $M=3$ for the Bernstein model for the baseline hazard functions, $\lambda_{0,k}(t), k = 1, 2$. For estimation, we generate 10,000 posterior samples after discarding the first 1,000 samples as burn-in. Trace plots suggest that the posterior sampling converges well.

The proposed method has three main components: the family-wise likelihood to handle missing genotypes, the ACJ likelihood to correct the ascertainment bias, and the frailty to capture the family-specific random effects. To evaluate the effects of these three components, we compare our approach with alternative approaches, under which there is (1) no missing genotype, (2) no ascertainment bias correction, and (3) no frailty. 

Table \ref{tb:sim} shows absolute biases and standard deviations of estimates under different approaches. 
For the baseline hazard $\lambda_{0,k}(t)$, bias and standard deviation are numerically integrated over $t$. We can see that the estimates without ascertainment bias correction are severely biased,  especially for $\beta_2$ and $\lambda_{0,2}(t)$, showing the importance of performing the ascertainment bias correction. In addition, the estimates with frailty tend to have smaller biases than those assuming no frailty. Lastly, the efficiency loss due to missing genotypes is generally small, suggesting that the family-wise likelihood efficiently utilizes the observed data. 

\begin{table}[!htbp]
\begin{center}
\begin{tabular}{cl rcrc c rcrc} \hline 
            &       &\multicolumn{4}{c}{No bias correction} && \multicolumn{4}{c}{Bias correction} \\ \cline{3-6}\cline{8-11}
		    &  Genotype     &\multicolumn{2}{c}{No fraility} & \multicolumn{2}{c}{Frailty} && \multicolumn{2}{c}{No fraility} & \multicolumn{2}{c}{Frailty} \\ \cline{1-6}\cline{8-11}
\multirow{2}*{$\beta_1$}
            & No missing &  1.1968 & (.3608) & 0.7818 & (.3385) && 1.0667 & (.3633) & 0.4905 & (.3420) \\
		    & Missing  &  1.4363 & (.4117) & 1.2627 & (.3421) && 1.2824 & (.4120) & 0.8681 & (.3942) \\ \cline{1-6} \cline{8-11}
\multirow{2}*{$\beta_2$}
            & No missing &  5.5993 & (.1973) & 4.7515 & (.2051) && 0.8659 & (.3220) & 0.2347 & (.4627) \\
		    & Missing  &  5.7480 & (.2225) & 5.4368 & (.2267) && 1.2012 & (.3016) & 0.4764 & (.3293) \\ \cline{1-6} \cline{8-11} 
\multirow{2}*{$\lambda_{0,1}(t)$}
            & No missing &  0.0227 & (.0366) & 0.0194 & (.0409) && 0.0190 & (.0394) & 0.0116 & (.0479) \\
		    & Missing  &  0.0184 & (.0374) & 0.0167 & (.0398) && 0.0166 & (.0395) & 0.0123 & (.0449) \\ \cline{1-6} \cline{8-11}
\multirow{2}*{$\lambda_{0,2}(t)$}
            & No missing &  0.1025 & (.0505) & 0.0914 & (.0518) && 0.0004 & (.0004) & 0.0043 & (.0038) \\
		    & Missing  &  0.1340 & (.0641) & 0.1116 & (.0540) && 0.0010 & (.0008) & 0.0053 & (.0045) \\ \hline
\end{tabular}
\caption{Absolute biases and standard deviations (in parentheses) of estimates based on 100 simulations. } \label{tb:sim}
\end{center}
\end{table}

\section{Application} \label{s::analysis} 
We apply the proposed methodology to analyze the LFS data. We consider three types of LFS-related cancers ($K=3$): breast cancer ($k=1$), sarcoma ($k=2$), and other cancers ($k=3$). Because the individuals with breast cancer in the LFS data are all female (Table \ref{tb::summary2}), we impose the following constraint on the hazard of developing breast cancer:
\begin{align} \label{eq:no.male.breast.model}
\lambda_1(t|G, X) =
\left\{
\begin{array}{ll}
0, & \mbox{for $X = 0$ (male)}, \\
\lambda_{0,1}(t) \xi_{1} \exp\{\beta_{G,1} G\},  & \mbox{for $X = 1$ (female)},
\end{array} \right.
\end{align}
while other types of cancer ($k = 2,3$) are assumed to follow the model of the form \eqref{eq::frailty.model}.
There is only one baseline covariate available in the LFS database (i.e., gender), however our method can readily accommodate more covariates. We ignore all cancers that occurred after 75 years of age and treat them as censored at age 75, since cancers diagnosed after 75 years of age are clinically irrelevant for estimating the penetrance of LFS.

\subsection{Model Parameter Estimates} \label{ss::para.est}
Posterior estimates for the regression coefficients $\bbeta_k$ and the inverse of the frailty variances $\bnu_k, k = 1, 2, 3$ are reported in Table \ref{tb::posterior}. Genotype has a strong effect on the incidence of all cancer types, with TP53 mutation carriers being more likely to have cancers. Gender also plays a significant role in sarcoma and other cancers. The regression coefficient of gender is negative, suggesting that males in this population are more likely to develop sarcoma and other cancers than females. 

The estimates of $\nu_k$s are quite large, which suggests that after accounting for the pedigree structure through the family-wise likelihood, within-family correlations are not very strong in this particular dataset. To check this, we compared the penetrance estimates obtained from our model to those from the model that does not include frailty and found them to be quite similar (see {\textit{Supplementary Materials}} Section D.3). Although the model without frailty may be preferred in practice due to its parsimony, we present the results of the frailty model to emphasize that our approach allows for further flexibility; the results are nearly identical in terms of the penetrance estimates. 

\begin{table}[!htbp]
\centering
\caption{Posterior estimates of regression coefficients $\bbeta$ and inverse variances of the frailty $\bnu$. 
} \label{tb::posterior}
\begin{tabular}{llrrrr} \hline
Cancer  & Parameter      &  Mean  &    SD &  2.5\% & 97.5\% \\ \hline
Breast  & Genotype  & 3.560  & 0.516 &  2.541 & 4.544  \\ \cline{2-6} 
        & (Frailty Var)$^{-1}$ &  6.126 & 1.850 & 3.185  & 10.347 \\ \hline
Sarcoma & Genotype    &  2.464 & 0.895 &  0.675 &  4.182 \\
        & Female      & -3.677 & 1.077 & -6.176 & -1.902 \\
        & Interaction &  0.971 & 0.548 & -0.110 &  2.040 \\ \cline{2-6} 
        & (Frailty Var)$^{-1}$&  6.574 & 1.990 & 3.490  & 11.227 \\ \hline
Others  & Genotype    &  1.576 & 0.769 &  0.072 &  3.072 \\
        & Female      & -0.993 & 0.186 & -1.366 & -0.647 \\
        & Interaction &  0.559 & 0.574 & -0.620 &  1.628 \\ \cline{2-6}
        & (Frailty Var)$^{-1}$ &  7.148 & 2.001 & 3.986  & 11.857 \\ \hline 

\end{tabular}
\end{table}

\begin{figure}[!t]
\centering
\subfigure[Breast]{
\includegraphics[width = .31\textwidth]{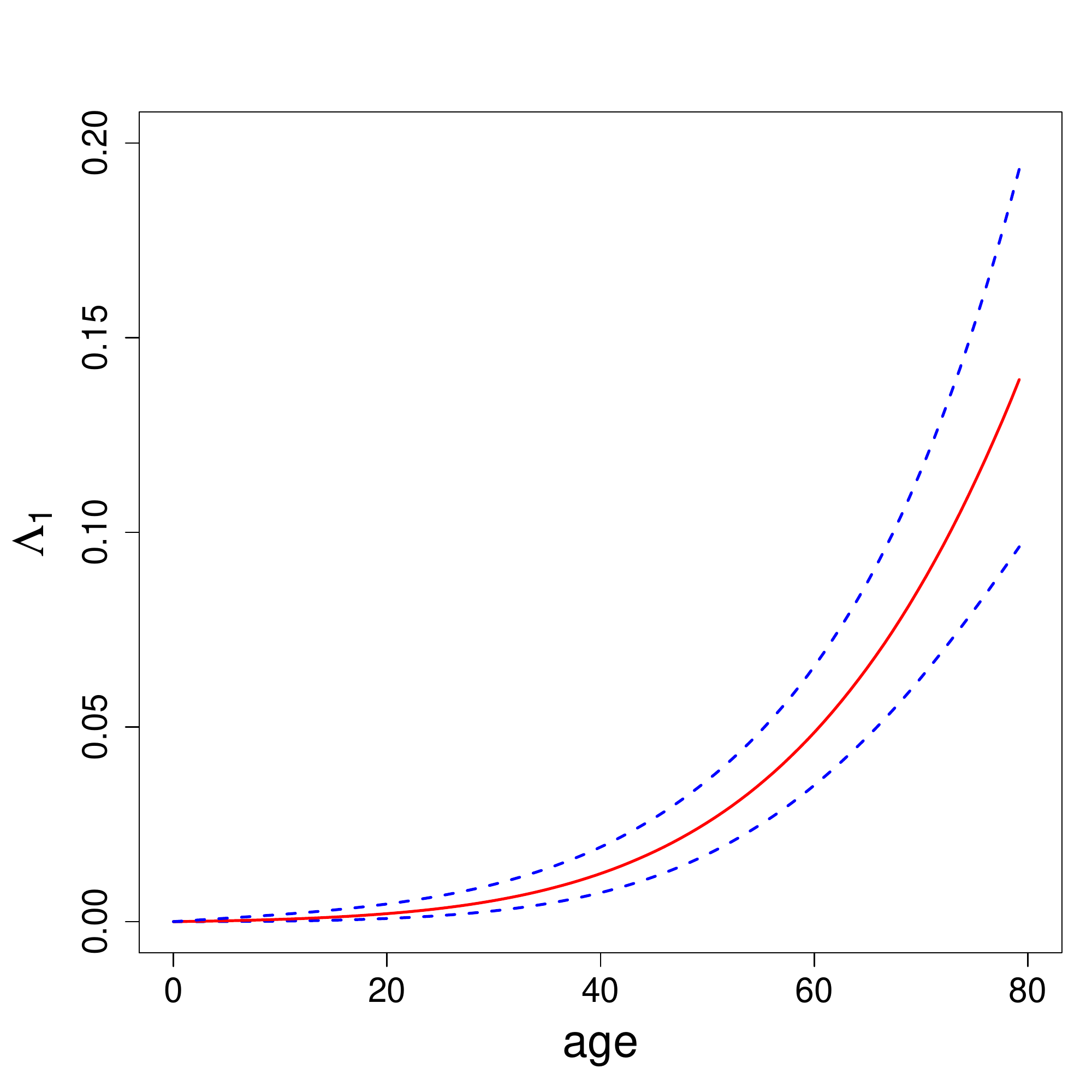}}
\subfigure[Sarcoma]{
\includegraphics[width = .31\textwidth]{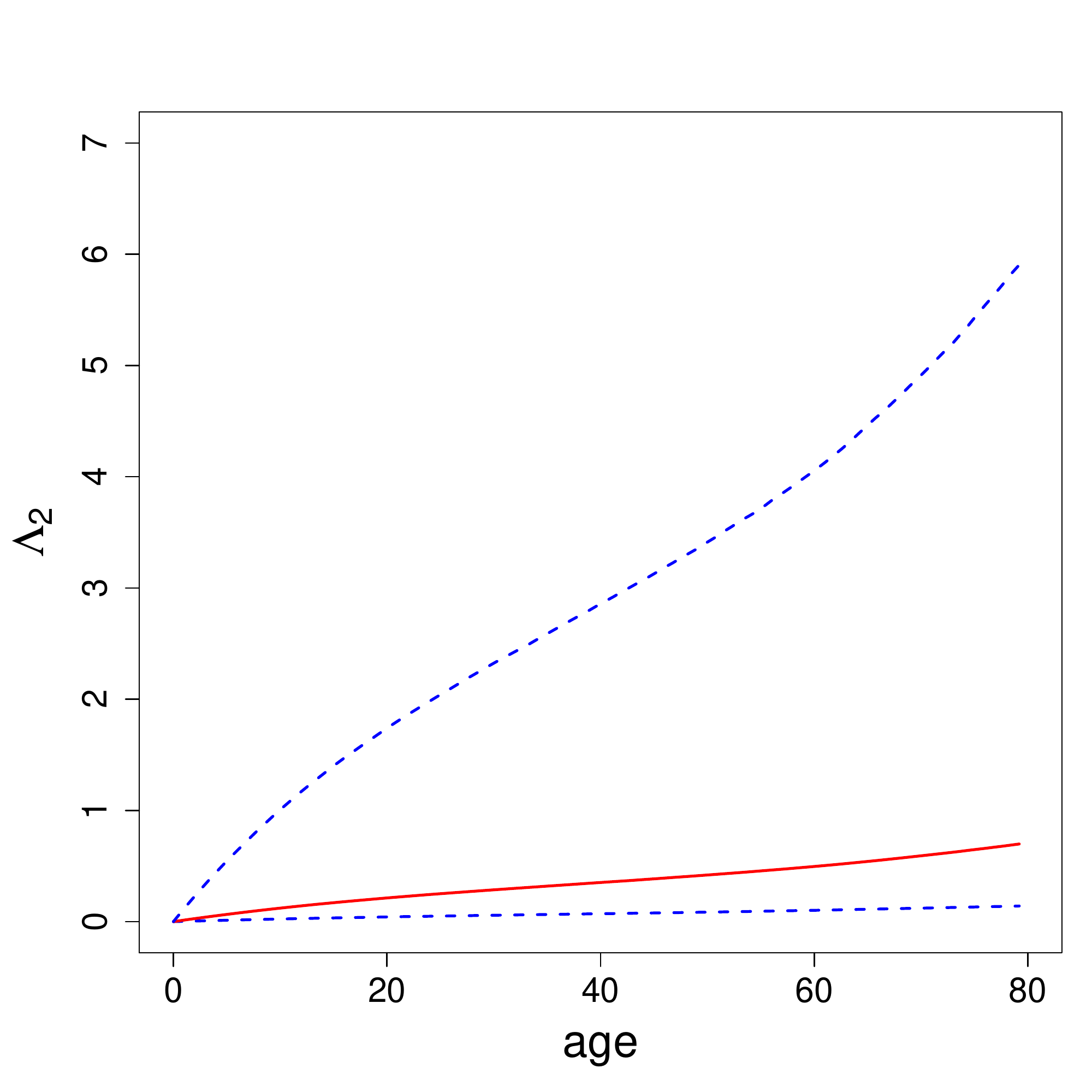}}
\subfigure[Others]{
\includegraphics[width = .31\textwidth]{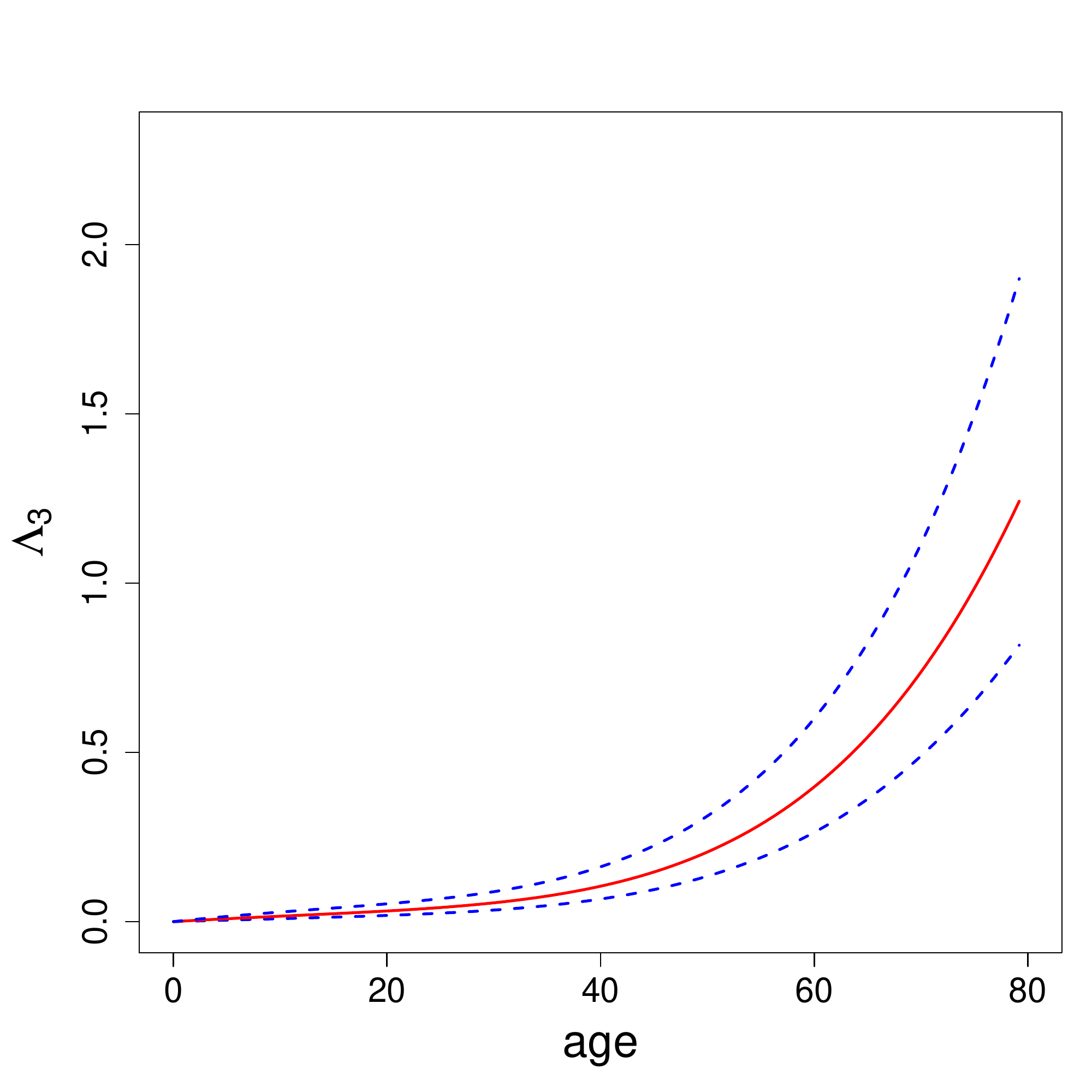}}
\caption{Posterior estimates of the cancer-specific cumulative baseline hazard functions for breast cancer (a), sarcroma (b) and other cancers (c). 
Dashed lines indicate 95\% credible band of the estimates.} \label{fg::baseline}
\end{figure}

Figure \ref{fg::baseline} depicts the posterior estimates of the cumulative baseline hazard.
Age has stronger effects on breast and other cancers than on sarcoma. The cumulative baseline hazards of breast and other cancers increase exponentially with age, while that of sarcoma increases approximately linearly with age. We observe that the uncertainty of the sarcoma baseline hazard estimate is much larger than those of the others. This is because the ascertainment bias is generated from the probands with sarcoma, which makes the ascertainment-bias-corrected likelihood \eqref{eq::acj} more sensitive to the parameters directly related to sarcoma.   

\subsection{Age-at-onset Penetrance}

\begin{figure}
\centerline{
\subfigure[Breast, $q_1(t|G,X)$]{
\includegraphics[width = 7cm, height = 7cm]{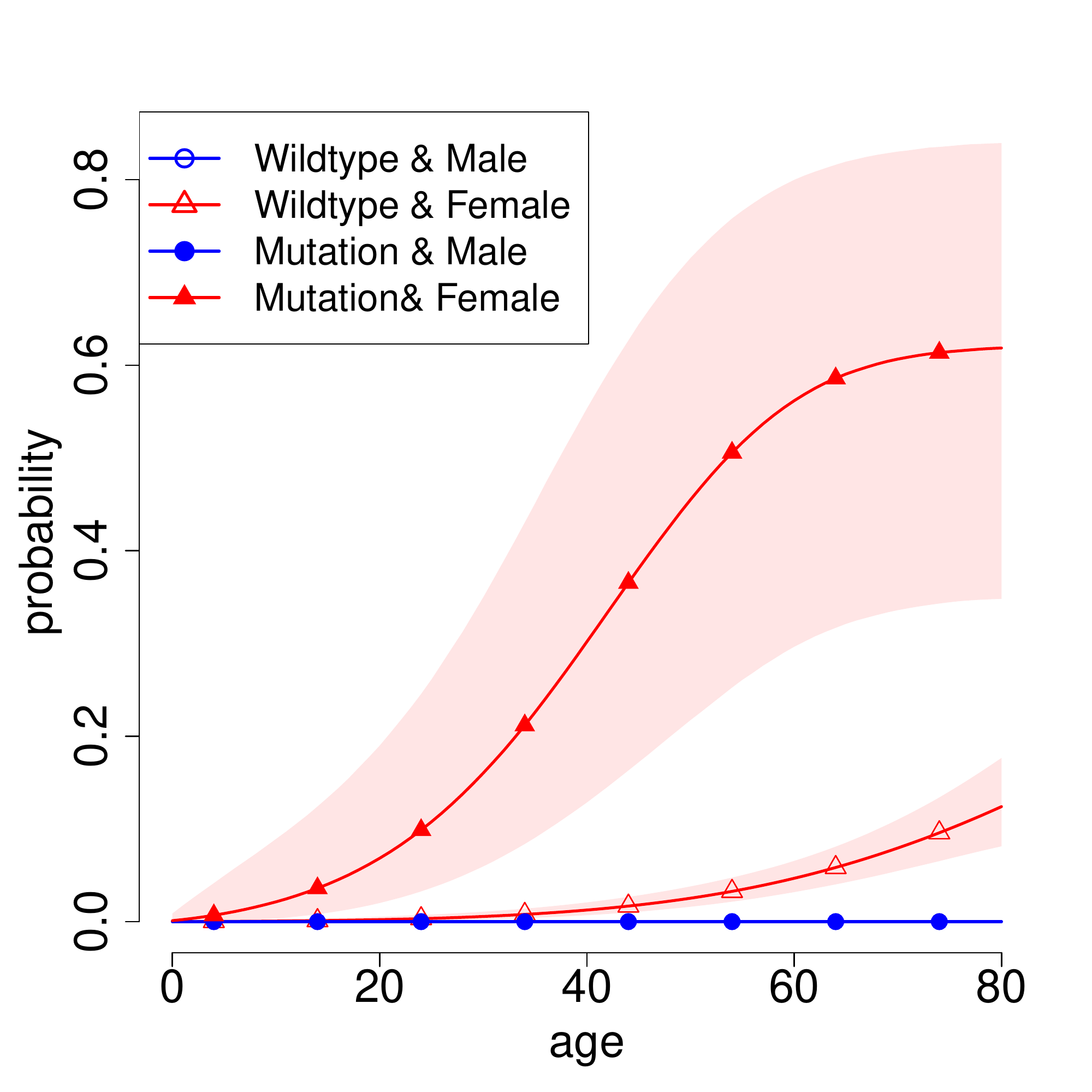}}
\subfigure[Sarcoma, $q_2(t|G,X)$]{
\includegraphics[width = 7cm, height = 7cm]{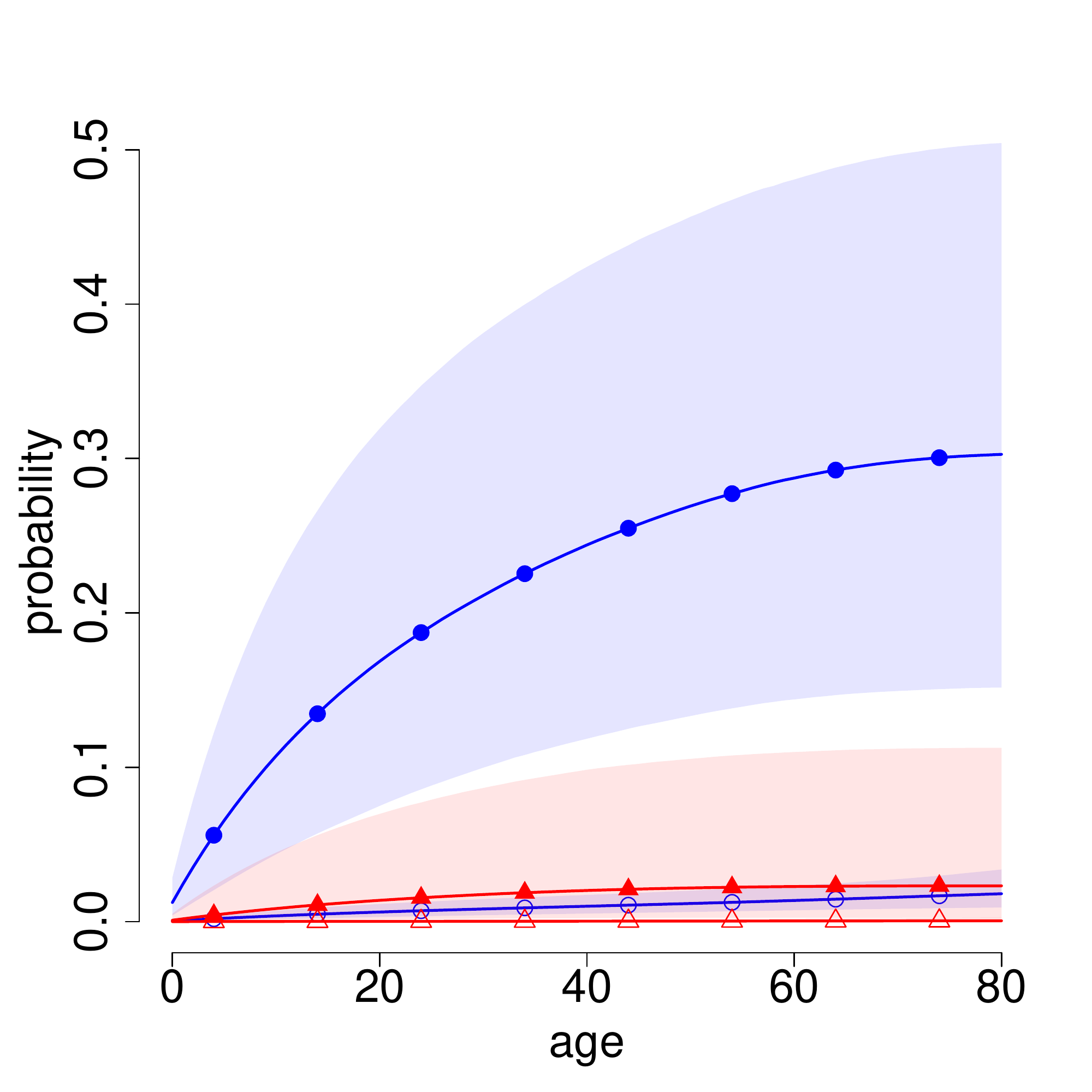}}}
\centerline{
\subfigure[Others, $q_3(t|G,X)$]{
\includegraphics[width = 7cm, height = 7cm]{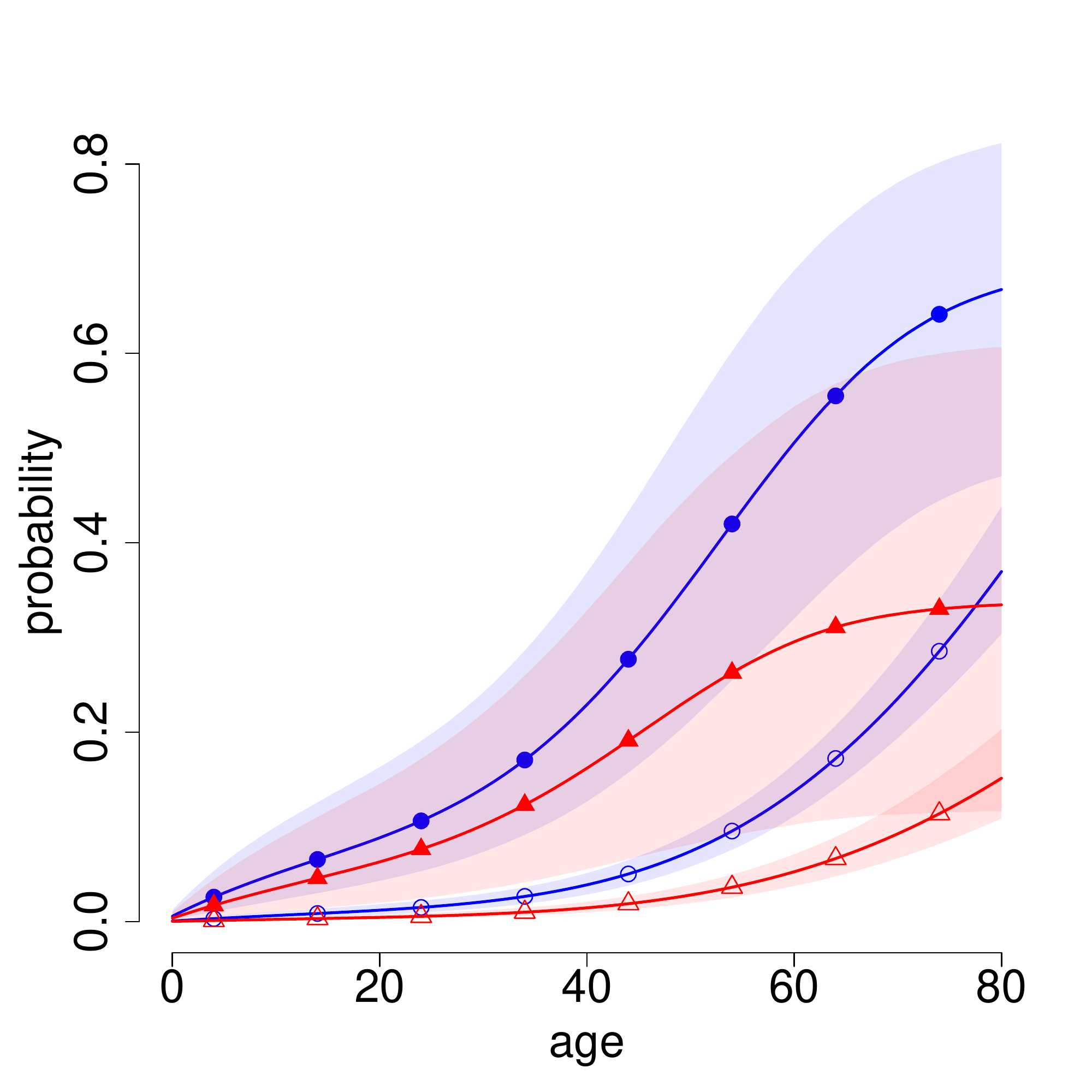}}
\subfigure[Overall, $q(t|G,X) = \sum_{k} q_k(t|G,X)$]{
\includegraphics[width = 7cm, height = 7cm]{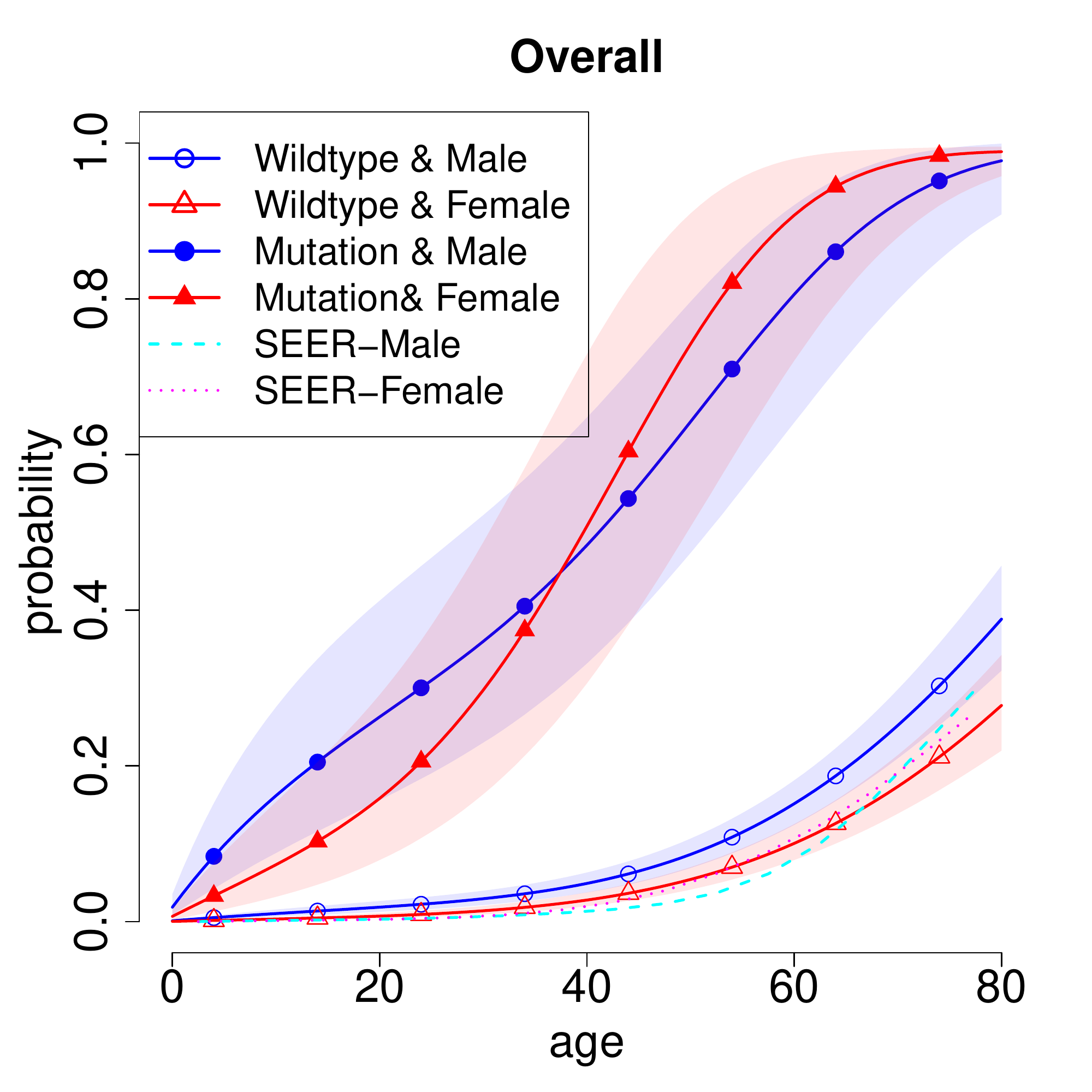}}}
\caption{Cancer-specific age-at-onset penetrances $q_k(t|G,X), k = 1, 2, 3,$ are depicted in (a)--(c), and the overall cancer penetrance $q(t|G, X)$ is given in (d). Shaded areas denote the 95\% credible bands. 
} \label{fg::survcurve}
\end{figure}

The first three panels (a)--(c) of Figure \ref{fg::survcurve} depict the estimated age-at-onset penetrances, $q_k(t|G, X), k = 1, \cdots, 3$, respectively, for breast cancer, sarcoma, and other cancers. It is not surprising that the TP53 mutation carriers ($G = 1$) have higher risk of developing cancer than the non-carriers ($G = 0$), regardless of cancer type. The patterns of cancer-specific penetrance are quite different across cancer types, which justifies the proposed cancer-specific approach. It is of clinical interest that there is a sizable chance that the female TP53 mutation carrier will develop breast cancer before 20 years of age, which is rarely seen in females with BRCA1 and BRCA2 mutations (two well-known susceptibility gene mutations for breast cancer) \citep{berry2002brcapro}. This suggests that early-onset breast cancer is an important feature of TP53 mutation.  We also find that non-carriers have very low probability of developing sarcoma, although the data contain many cases of sarcoma in non-carriers due to the use of individuals with sarcoma as probands for collecting the samples (see Table \ref{tb::summary2}). In contrast, ignoring the ascertainment bias leads to substantially biased estimates, see {\textit{Supplementary Materials} Section D.2} for the comparison between our estimates and the estimates without performing ascertainment bias correction.

Figure \ref{fg::survcurve}, panel (d) shows the overall age-at-onset penetrance obtained by stacking three cancer-specific penetrances, i.e., $q(t|G,X)= \sum_{k=1}^3 q_k(t|G, X)$. The overall age-at-onset penetrance quantifies the probability of having any type of cancer by a certain age for carriers of TP53 mutations. Among the non-carriers, females have lower cancer risk than males; whereas the female mutation carrier has higher risk than the male mutation carrier due to the excessively high risk of the female carrier developing breast cancer. Overall, TP53 mutation  carriers have very high lifetime risk of developing cancer, demonstrating the importance of the accurate detection of TP53 germline mutations.

\subsection{Personalized Risk Prediction} \label{s::risk.prediction}
An important application of our analysis results and estimate of age-at-onset penetrance $q_k(t|G, X)$ is to provide a personalized risk prediction for future subjects who are at risk of developing LFS-related cancers. Our prediction method has two important advantages. First, it allows us to incorporate the subject's family cancer history to make more accurate risk prediction. Second, it is capable to make risk prediction for a subject without knowing his/her genotype. This 
is desirable because in practice, genetic test is often of a great financial and psychological burden for patients. Making risk prediction without performing a genetic test allows us to 
quickly detect individuals with high risk of LFS and provide prompt and proper clinical treatments during an early stage of disease, which is particularly important in the management of rare diseases such as LFS. Specifically, given a family's cancer history $\bH_i$ and covariates $\bX_i$, the risk that the $j$th individual in the $i$th family will develop the $k$th type of cancer by age $t$, $R_{ijk}(t|\bH_i, \bX_i)$, is predicted by
\begin{align} \label{eq:cs.risk}
R_{ijk}(t|\bH_i, \bX_i) = \Pr(T_{ij} \le t, D_{ij} = k|\bH_i, \bX_i) = \sum_{G_{ij} \in \{0, 1\}} \Prob(G_{ij}|\bH_i, \bX_i) q_k(t|G_{ij}, X_{ij}).
\end{align}
That is, the predicted cancer-specific risk is a weighted average of the cancer-specific penetrance $q_k(t|G_{ij}, X_{ij})$. The weight $\Prob(G_{ij}|\bH_i, \bX_i)$, also known as  carrier probability,  is the likelihood that the subject carries a specific genotype $G_{ij}$, given his/her family cancer history $\bH_i$ and covariates $\bX_i$. It can be routinely calculated using Bayes' rule and Mendelian laws of inheritance, see {{\it Supplementary Materials} Section B} for details. As we assume that the subject's genotype $G_{ij}$ is unknown, the calculation of the risk in  (\ref{eq:cs.risk}) is marginalized over all possible values of $G_{ij}$. 

\begin{figure}[!htbp]
\centering
\subfigure[Family 1]{
\includegraphics[width = 0.45\textwidth]{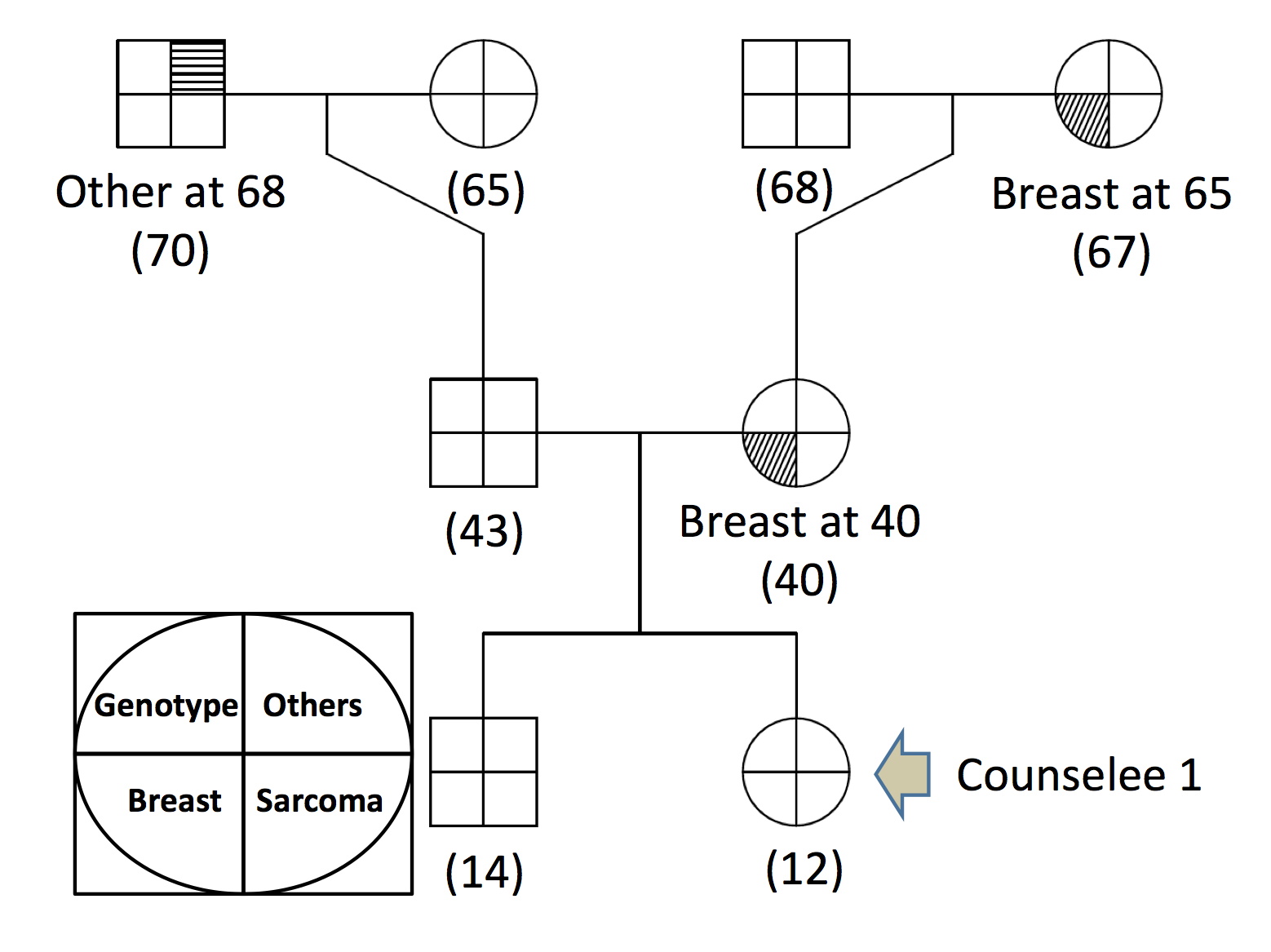}}
\subfigure[Family 2]{
\includegraphics[width = 0.45\textwidth]{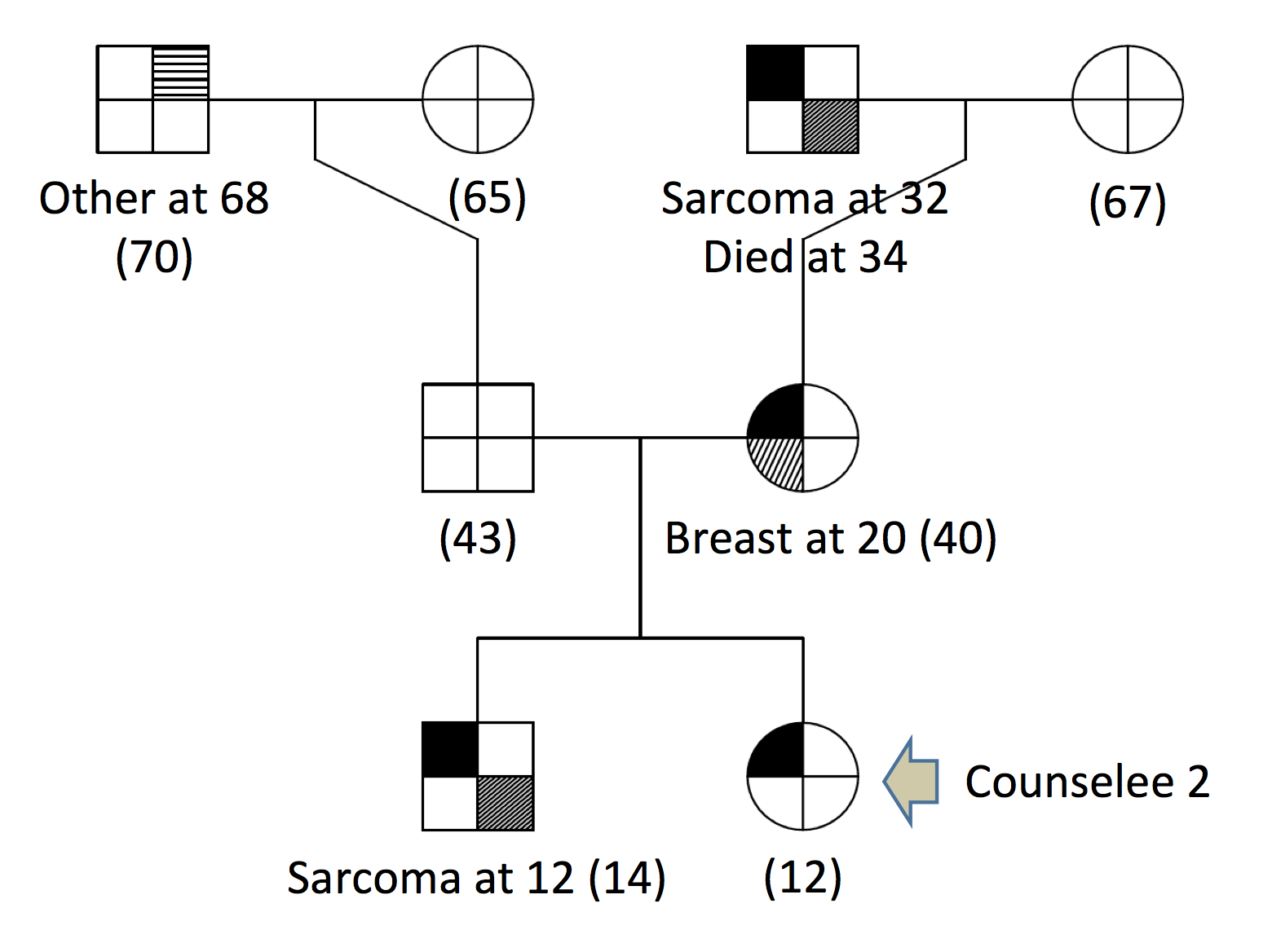}}
\caption{Pedigrees of two families, where square and circle represent male and female subjects, respectively. The symbol is partitioned into four sections, which represent statuses of genotype (topleft), breast cancer (bottom left), sarcoma (bottom right), and other cancers (topright). Filled sections represent that the subject carries a mutated allele or had a certain type of cancer.  The number in the parentheses is subject's current age. } \label{fg::illu_ped}
\end{figure}

\begin{figure}[!htbp]
\centering
\subfigure[Counselee 1 ($G$ is known)]{
\includegraphics[width = 0.47\textwidth]{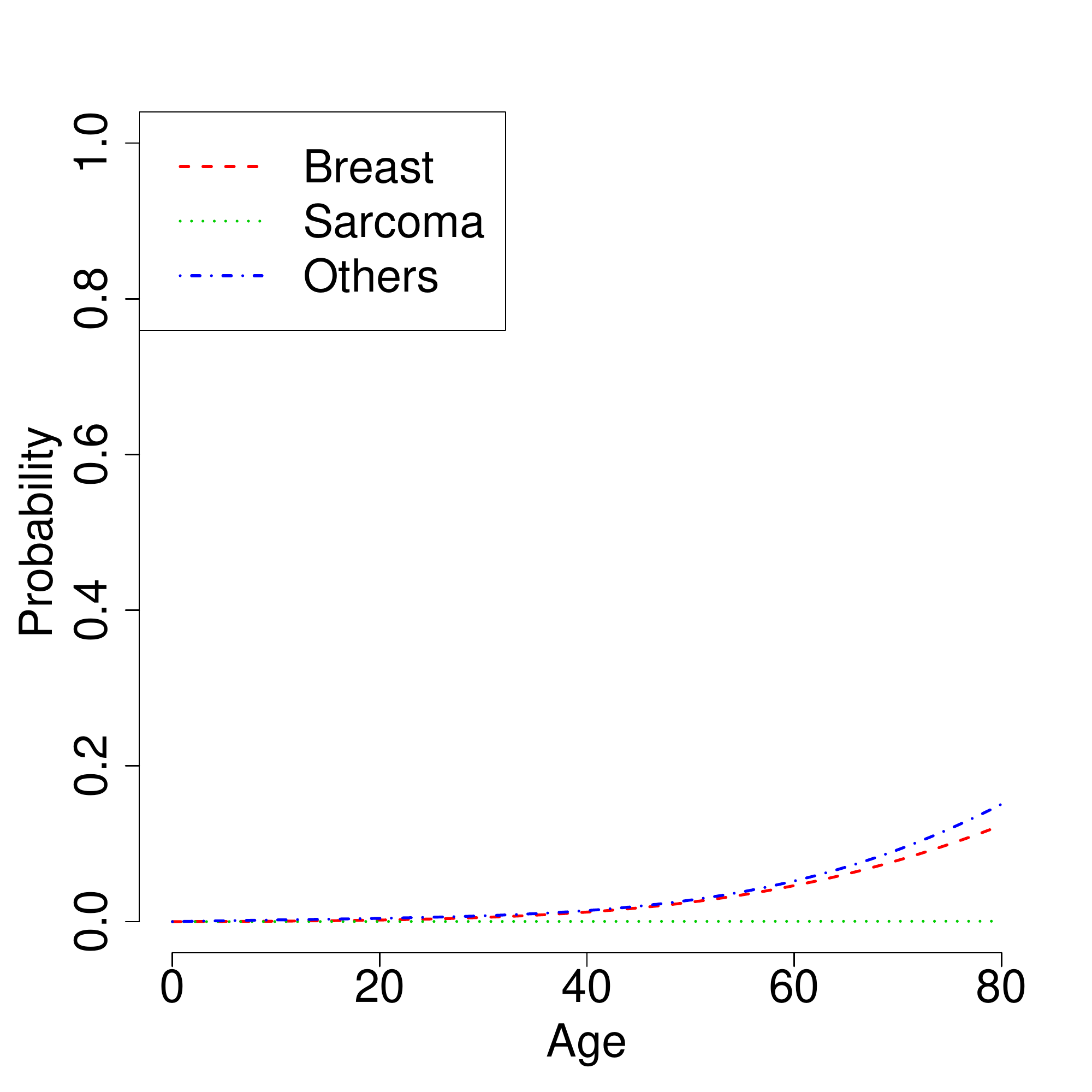}}
\subfigure[Counselee 1 ($G$ is unknown)]{
\includegraphics[width = 0.47\textwidth]{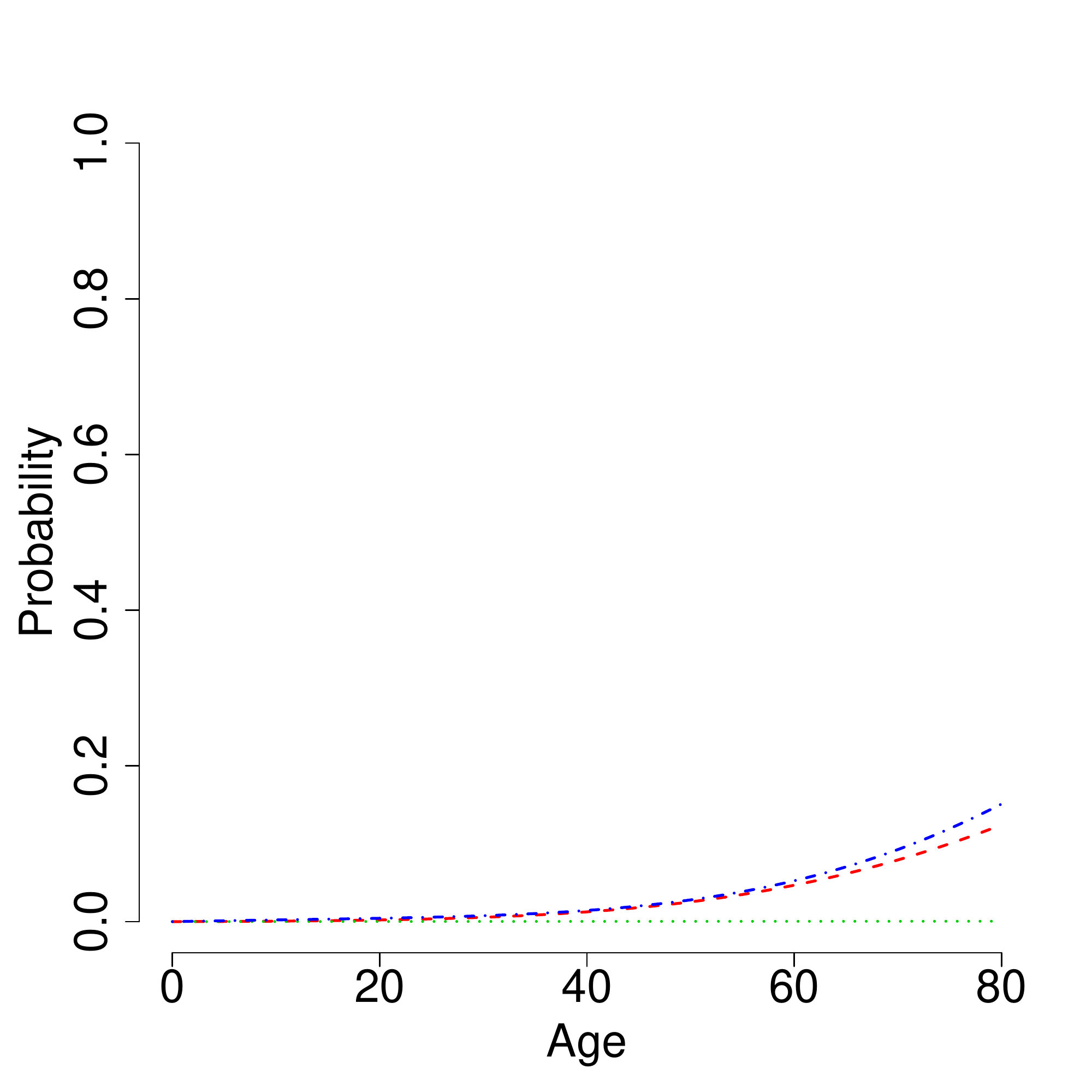}}
\subfigure[Counselee 2 ($G$ is known)]{
\includegraphics[width = 0.47\textwidth]{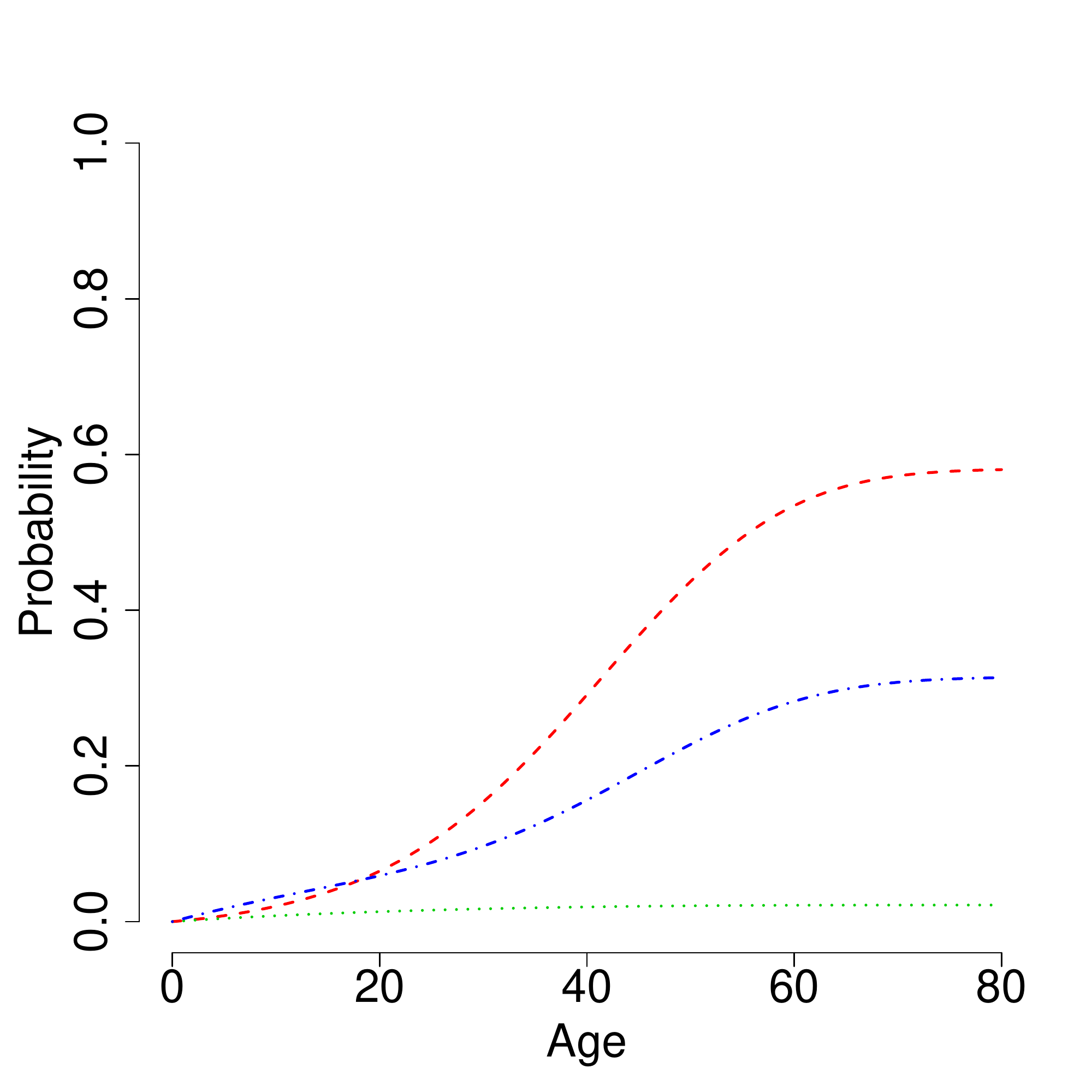}}
\subfigure[Counselee 2 ($G$ is unknown)]{
\includegraphics[width = 0.47\textwidth]{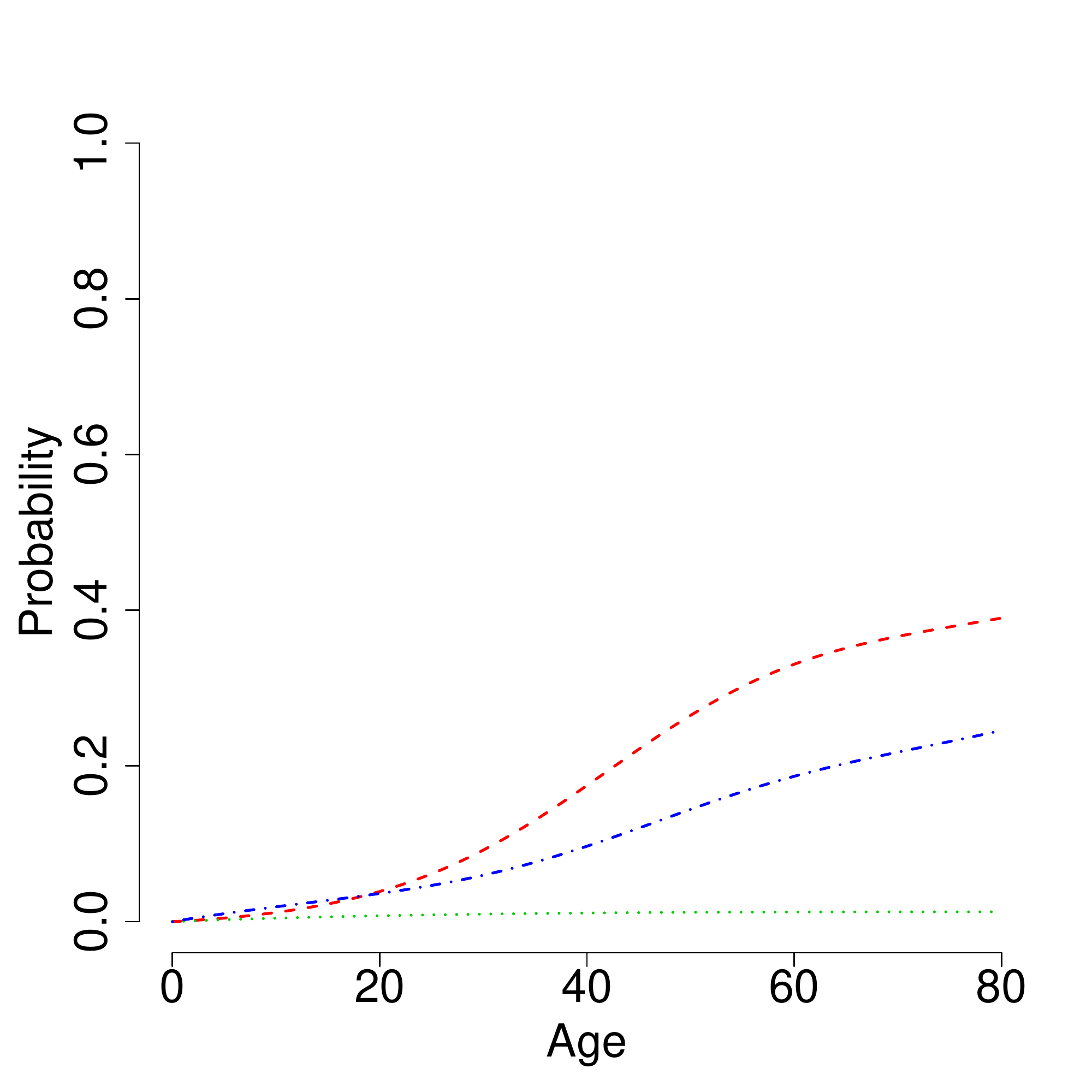}}
\caption{Predicted cancer-specific risk for counselees 1 and 2 when their genotypes $G$ are known or unknown.} \label{fg::illu_risk}
\end{figure}

To illustrate the utility of our method, consider two hypothetical families 
that have similar pedigree structures, but  different genotypes and cancer histories, as shown in Figure \ref{fg::illu_ped}. Family 1 does not carry the mutated allele and has three cases of cancer (two breast and one other cancers), and family 2 carries the mutated allele with four cases of cancer (one breast, two sarcoma and one other cancers).
As mothers (the second generation) in both families had breast cancer, it is of great interest to predict the cancer risk for their daughters, referred to as counselees 1 and 2 in Figure \ref{fg::illu_ped}. We consider two situations: the genotypes of the counselees are known or unknown. Specifically, when
 the genotypes of the counselees and their family are unknown, we predict the cancer risk for the counselees based on  equation \eqref{eq:cs.risk} with the cancer-specific penetrance estimated from the LFS data. When the genotypes of the conselees are known (i.e., conselee 1 is non-carrier and 2 is carrier), the risk prediction is straightforward and the cancer risk of the conselees is simply the estimated cancer-specific penetrances $q_k(t|G, X)$. Figure \ref{fg::illu_risk} shows the predicted cancer-specific risks of the counselees when their genotypes are known and unknown. Clearly, counselee 2 has a substantially higher risk of developing cancer than the counselee 1. Based on this result, we may recommend more frequent cancer screening for counselee 2. We note that counselee 2 has a very low risk of developing sarcoma although her family has two cases of sarcoma. This is because, as shown in Figure \ref{fg::survcurve}(b), the penetrance for sarcoma is high in male, but very low in female.
 
\subsection{External and Interval Validation}
As an external validation, we compare our estimates of non-carrier penetrance to those provided by the National Cancer Institute on the basis of the Surveillance, Epidemiology, and End Results (SEER) data. SEER is an authoritative source of information on cancer incidence and survival in the United States. It currently collects and publishes cancer incidence and survival data from population-based cancer registries that cover approximately 28\% of the US population. SEER is the only comprehensive source of population-based information in the United States that includes the stage of cancer at the time of diagnosis and patient survival data. The SEER estimate can be regarded as a reference estimate for the normal US population (i.e., non-carrier). More details regarding SEER estimates can be found at \url{http://seer.cancer.gov}.

\begin{figure} [!htbp]
\centering
\subfigure[Breast (female)]{
\includegraphics[width = .31\textwidth]{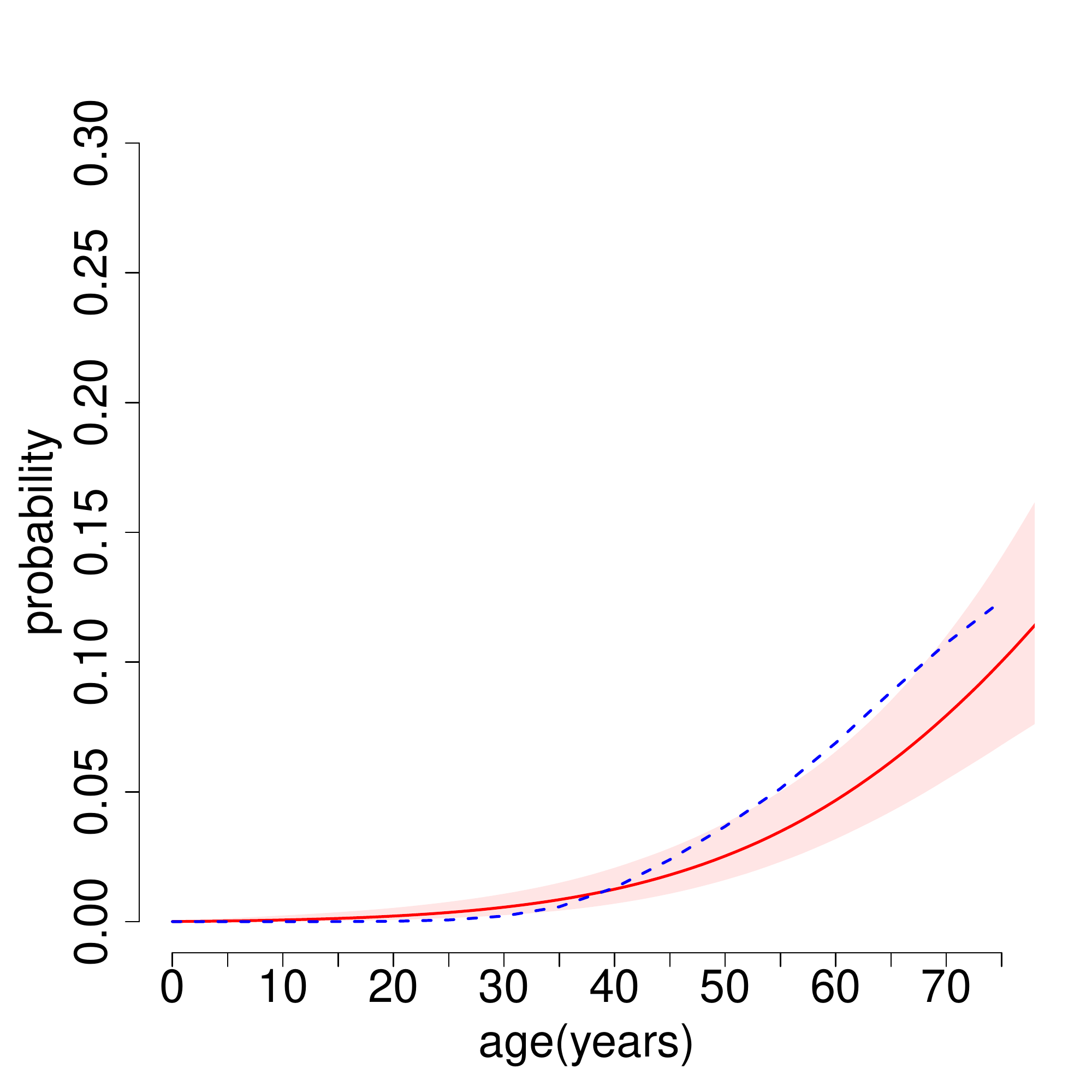}}
\subfigure[Sarcoma]{
\includegraphics[width = .31\textwidth]{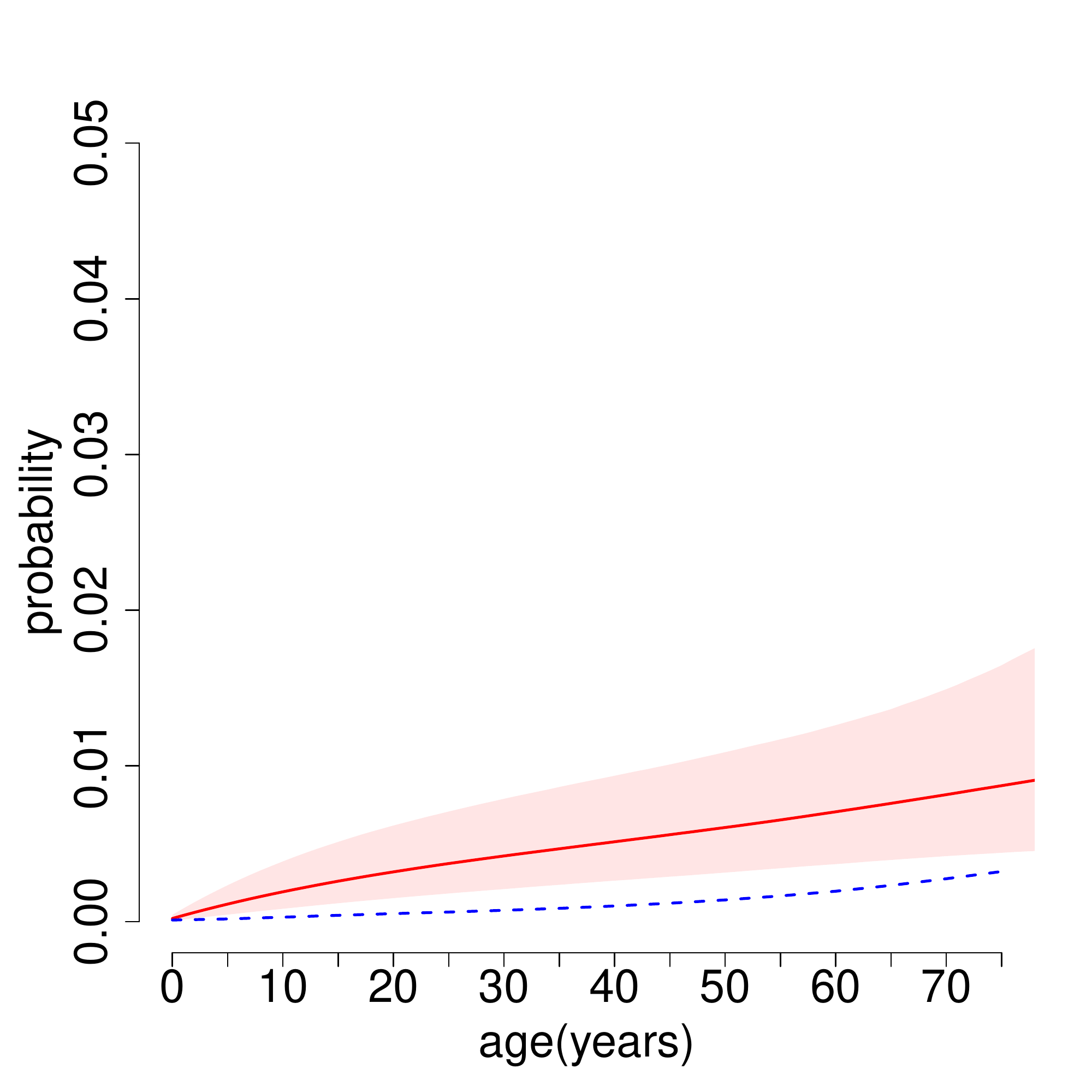}}
\subfigure[Overall]{
\includegraphics[width = .31\textwidth]{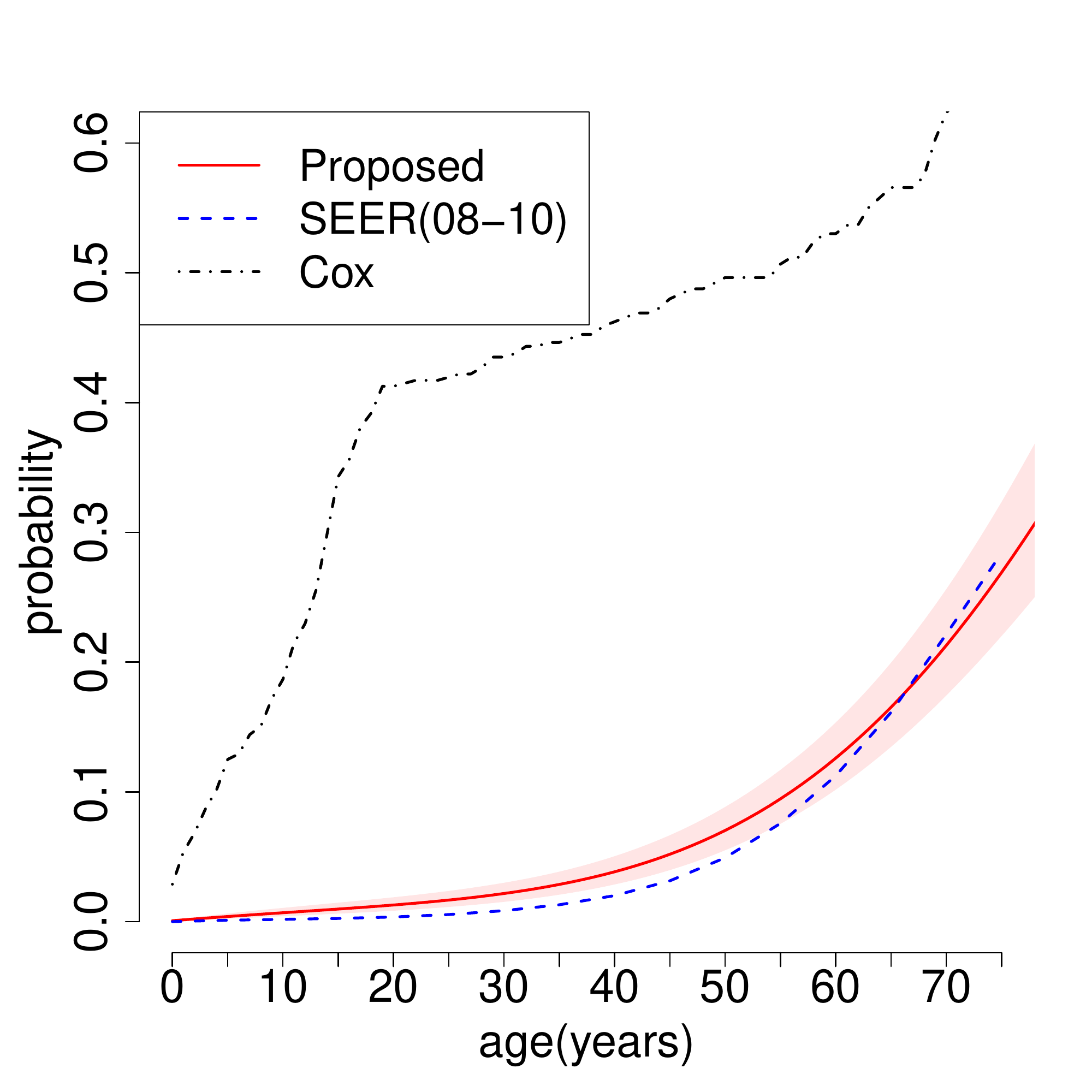}}
\caption{External validation of the estimated penetrance for non-carrier (solid lines) through comparison to the SEER estimates (dashed lines). Shaded areas represent the 95\% credible bands. For panel (c), the dotted line indicates the estimate based on the Cox model.
} \label{fg::ex.validation}
\end{figure}

Figure \ref{fg::ex.validation} compares the penetrance of breast cancer, sarcoma, and all cancers for non-carriers to the most recent SEER estimates based on the data collected from 2008 to 2010. We can see that the estimates of non-carrier penetrance are generally consistent with the corresponding SEER estimates, suggesting that the proposed methodology performs well. For the purpose of comparison, we also show the estimate of the overall cancer penetrance based on the conventional Cox model for the time to cancer diagnosis using subjects with known genotypes. As shown in Figure \ref{fg::ex.validation}, panel (c), the estimate of the overall cancer risk based on the proposed method is much closer to the SEER estimate than the estimate based on the Cox model.

\begin{figure} [!h]
\begin{center}
\centerline{\includegraphics[width = 0.7\textwidth]{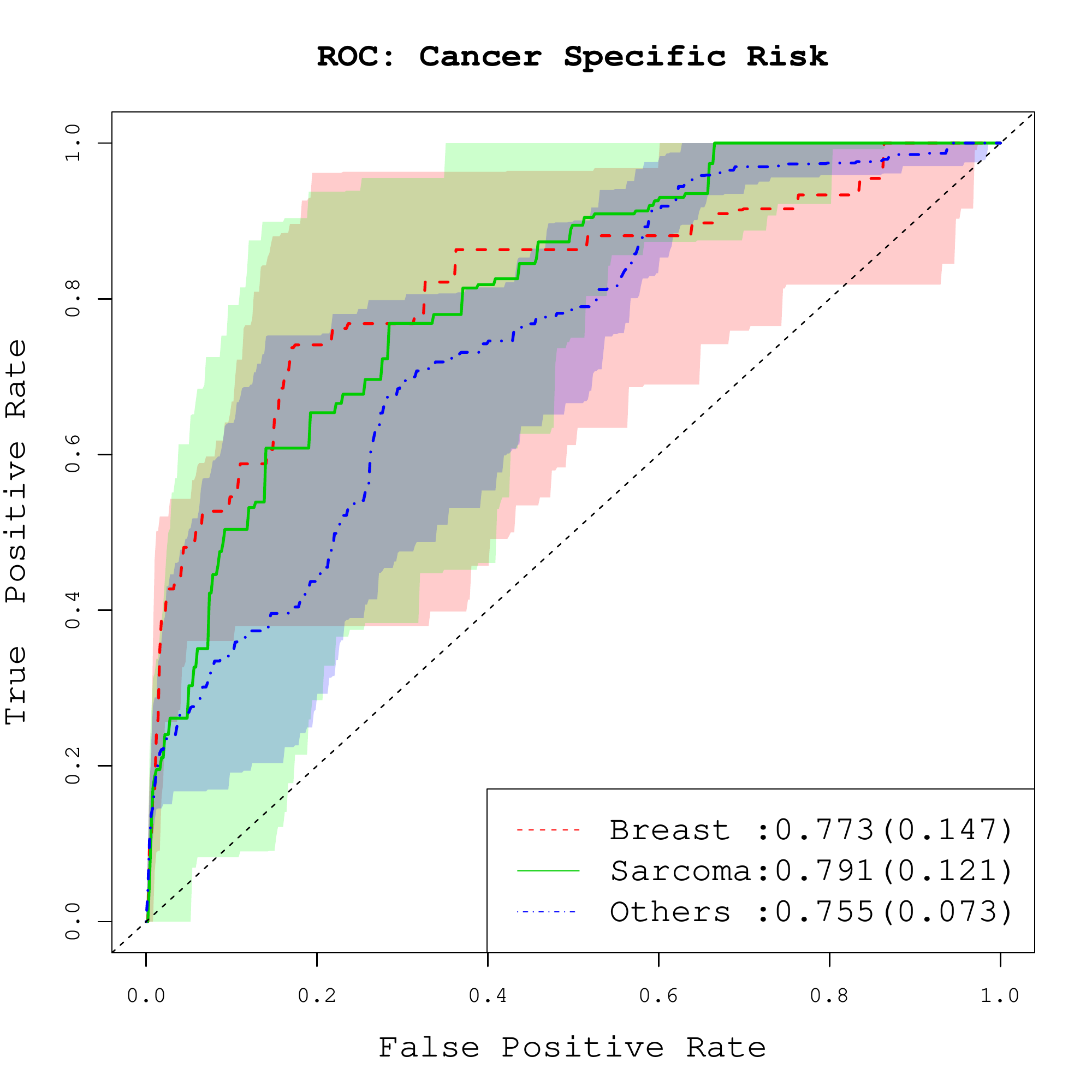}}
\caption{The ROC curve for the cancer-specific risk prediction at age 50. Values in the parentheses are standard deviations of the AUC. Shaded areas represent pointwise 95\% variations of the ROC curves from different random partitions.}
\label{fg::cv.roc}
\end{center}
\end{figure} 

We conduct internal validation through cross-validation. First, we randomly split the data (i.e., 186 families) into two halves. We use one half (i.e., 93 families) as the training families $\{(\bH_i^{\tr}, \bX_i^{\tr}, \bG_{obs,i}^{\tr}), i = 1, \cdots, 93\}$, and the other half as the test families, $\{(\bH_{i^\prime}^{\tr}, \bX_{i^\prime}^{\tr}, \bG_{obs,i^\prime}^{\tr}),$ $i^\prime = 1, \cdots, 93\}$. Next, we estimate the cancer-specific penetrance using the training families, denoted by $\hat q_k^{\tr}(t|G, X)$. Based on this estimate and equation \eqref{eq:cs.risk}, we predict the cancer-specific risk at a given age $t_c$ for subjects in the test families, i.e., $R_{i^\prime jk}(t_c| \bH_{i^\prime}^\ts, \bX_{i^\prime}^\ts)$. Given a certain risk cutoff $\psi$, we predict that a subject will have $k$th type of cancers by age $t_c$ if $R_{i^\prime jk}(t_c| \bH_{i^\prime}^\ts, \bX_{i^\prime}^\ts)>\psi$. By varying the risk cutoff $\psi$ and comparing the predicted cancer status with the actually observed cancer status of the test families, we obtain the receiver operating characteristic (ROC) curves of our cancer risk prediction model. Figure \ref{fg::cv.roc} depicts the ROC curve of the predicted risk of the test family members at age 50 years for different cancer types. These results show reasonable performance, with the area under the ROC curves (AUC) being 0.773, 0.791 and 0.755 for predicting breast cancer, sarcoma and other cancers, respectively. For breast cancer, the ROC curves are generated from the females only since we assume no breast cancer for the males. We also consider the ROC curves for other caner-onset ages, $t_c = 30, 40$, and $60$ years. The results are generally similar to  that of $t_c=50$ years, see {\textit{Supplementary Materials} Section F}.

\subsection{Model Comparison} \label{ss:compare}

Due to the complicated structure of the LFS data (e.g., family structure, missing genotype, ascertainment bias and competing risks), standard model diagnosis tools for survival models, such as residuals \citep{schoenfeld1982partial,therneau1990martingale} and chi-squared goodness-of-fit tests \citep{hjort1990goodness,hollander1992chi,li1993generalized}, are not applicable here. We assess the adequacy of the proposed model through model comparison. 
We consider four alternative models. The first three models are obtained by replacing the Bayesian nonparametric baseline hazard model with three parametric models: the exponential, Weibull, and piecewise-constant models, respectively. {For the piecewise-constant model, we use four equally spaced knots to obtain five partitions.}  The fourth model is obtained by removing the frailty $\xi_{i, k}$ from the competing risk model (\ref{eq::frailty.model}). We use two metrics to measure the goodness of fit of the models: the deviance information criterion (DIC) and conditional predictive ordinate \citep[CPO,][]{ibrahim2005bayesian}. The DIC measures the overall goodness of fit of a model and the CPO measures the predictive ability of a model. 
The CPO for the $i$th family is defined as 
\begin{align} \label{eq::cpo}
\mbox{CPO}_i = \Pr(\bH_i|{\mathcal{D}_{(-i)}}, \bG_{i, obs}, \bX_i, \bxi_i, {\mathcal{A}_i}) = \
\left[ E \left(\frac{1}{\Pr(\bH_i|\bG_{i, obs}, \bX_i, \bxi_i, \btheta, {\mathcal{A}_i})} \right) \right]^{-1}
\end{align}
where $\mathcal{D}_{(-i)} = (\bH_{(-i)}, \bG_{(-i), obs}, \bX_{(-i)})$ represents the data with the $i$th family data deleted, and the expectation is made with respect to the posterior distribution of $\btheta$.  The Monte Carlo approximation of \eqref{eq::cpo} is given by
$$
\widehat{\mbox{CPO}}_i = \left[\frac{1}{L}\sum_{\ell = 1}^L   \frac{1}{\Pr(\bH_i|\bG_{i, obs}, \bX_i, {\hat \bxi_{i,(l)}}, {\hat \btheta_{(\ell)}}, {\mathcal{A}_i})}\right]^{-1}
$$
where $\hat \bxi_{i,(\ell)}$ and $\hat \btheta_{(\ell)}$ denote posterior samples from the $\ell$th MCMC iteration, $\ell=1, \cdots, L$.

Table \ref{tb::compare} shows the DIC and $\sum_{i=1}^I \log \widehat{\mbox{CPO}}_i$, known as the pseudo-marginal log-likelihood (PsML), for the different models. Smaller DIC values and larger PsML values suggest a better model. The proposed model based on Bernstein polynomials provides better goodness of fit and predictive ability than the models with exponential, Weibull, or piecewise-constant baseline hazards. The difference between the proposed model and the model without frailty is small, suggesting a weak within-family correlation.
 This is concordant with our finding that $\bnu$ estimates are large (see Table \ref{tb::posterior}). For the purpose of comparison, we also perform the analysis based only on the subset of the data for whom the genotypes are observed, and the analysis without ascertainment bias correction.  The estimates of cancer-specific penetrance under different approaches are provided in {\it Supplementary Materials} (Section D). 

\begin{table}[!htbp]
\begin{center}
\begin{tabular} {ccccc} \hline
         & Baseline         &  Frailty   &        &          \\
   Model & hazard           &  included  &    DIC &.   PsML  \\ \hline
       1 &Exponential       & Yes        & 3273.7 &  {$-$1657.120} \\ 
       2 & Weibull          & Yes        & 3020.2 &  {$-$1512.252} \\ 
       3 & Piecewise        & Yes        & 3010.3 &  {$-$1513.405} \\ 
       4 & Bernstein.       & No         & 2989.3 &  {$-$1499.735} \\ 
Proposed & Bernstein        & Yes        & 2983.7 &  {$-$1499.689} \\ \hline
\end{tabular}
\caption{Comparison of the proposed model with four alternative models.     } \label{tb::compare}
\end{center}
\end{table}

\subsection{Sensitivity Analysis} \label{ss::sensitivity}
We consider nine different combinations of priors for $\gamma_{m,k}$ and $\nu_k$: three different priors for $\gamma_{m,k}$ including flat prior, $Gamma(0.01, 0.01)$, and $Gamma(1, 1)$; and three priors for $\nu_k \sim Gamma(0.01, 0.01), Gamma(0.1, 0.1)$ and $Gamma(1, 1)$. The results (see {\textit{Supplementary Materials}} Section E) show that the estimates are not particularly sensitive to the choice of priors. 

\section{Discussion} \label{s::discussion}
In the LFS study, estimating cancer-specific penetrance is not trivial under the presence of competing risks, but is essential for providing better treatment that is personalized to the patient's needs. We developed a cancer-specific age-at-onset penetrance model and proposed an associated Bayesian estimation scheme. The proposed method can incorporate all the family histories in the estimation by exploiting the family-wise likelihood. We also corrected the ascertainment bias, which is an important task in family data studies of rare diseases.

One detriment when modeling the cause-specific hazard in competing risk analysis is that covariate effects on the subdistribution (i.e., cancer-specific penetrance) are not interpretable. As an alternative, \citet{fine1999proportional} proposed a proportional model for the subdistribution that enables us to directly assess the covariate effects on the corresponding cancer-specific penetrance. It is not difficult to equivalently rewrite the individual likelihood in terms of the cancer-specific penetrance and the associated derivative \citep{maller2002analysis}. The family-wise likelihood approach can be similarly applied to this alternative modeling approach.

In the LFS study, a patient can have multiple primary cancers during his or her lifetime. In the current approach, we consider only the first cancer that occurred and discard all the subsequent cancer history. In order to incorporate a longitudinal history that may involve multiple cancers, our approach can be extended to the so-called multi-state model \citep{putter2007tutorial} to recurrently observe multiple failures. In theory, the multi-state model can be regarded as an extended version of the competing risk model. However, it is practically challenging to collect data for a sufficient number of subjects who have multiple primary cancers in order to attain an appropriate level of estimation accuracy.

\section*{Supplementary Material}
\label{SM}
{\it Supplementary Material} includes an illustrative example of the peeling algorithm, a description of the carrier probability estimation based on family cancer history, additional simulation results for different baseline hazard models, penetrance of LFS estimated by various competing methods, prior sensitivity analysis, and cross-validated ROC curves at different ages. 

\bibliographystyle{dcu}
\bibliography{references}

\end{document}